%% file: oc_covid_paper.tex
\documentclass[12pt]{article}
\usepackage[top=1.2in, bottom=1.2in, left=1in, right=1in]{geometry}
\usepackage{graphicx}
\usepackage{amsthm}
\usepackage{amsmath}
\usepackage{amsfonts}
\usepackage{amssymb}
\usepackage{mathtools}
\usepackage{pdfpages}
\usepackage{blkarray} % for labels on matrix columns and rows
\usepackage{natbib}
\usepackage{fancyhdr}
\usepackage{footmisc}
\usepackage[english]{babel}
\usepackage{txfonts}
\usepackage{textcomp, gensymb} %for degrees C
\usepackage{caption}
\usepackage{listings,relsize}
\usepackage{setspace}
\usepackage{tcolorbox}
\usepackage{chngcntr}
\usepackage{ulem}
\usepackage{lineno}
\usepackage{booktabs}
\usepackage{url}
\usepackage{makecell}
\usepackage{dsfont}
\usepackage{xcolor}
\usepackage{tabularx}

\graphicspath{{figures/}} %Setting the graphicspath
\usepackage[colorinlistoftodos]{todonotes}
\usepackage{tikz}
\usetikzlibrary{arrows}

\tikzset{ shorten <>/.style={ shorten >=#1, shorten <=#1 } }

\input{custom_commands.tex}

\begin{document}

% % Title and Authors

\begin{center}
{\Large Semi-parametric modeling of SARS-CoV-2 transmission using tests, cases, deaths, and seroprevalence data}\\ \ \\

Damon Bayer$^{1}$, Isaac H. Goldstein$^{1}$, Jonathan Fintzi$^{2}$,
Keith Lumbard$^3$, Emily Ricotta$^4$, Sarah Warner$ ^5 $, Lindsay M. Busch$ ^6 $, Jeffrey R. Strich$ ^5 $, Daniel S. Chertow$^5$, Daniel M. Parker$^7$, Bernadette Boden-Albala$^7$, Alissa Dratch$^8$, Richard Chhuon$^8$, Nichole Quick$^9$, Matthew Zahn$^8$, and Volodymyr M. Minin$^{1,10,\dagger}$\\

\footnotesize{$^{1}$Department of Statistics, University of California, Irvine, California, U.S.A.}\\
\footnotesize{$^2$Biostatistics Research Branch, National Institute of Allergy and Infectious Diseases, Rockville, Maryland, U.S.A.}\\
\footnotesize{$^3$Clinical Monitoring Research Program Directorate, Frederick National Laboratory for Cancer Research, Frederick, Maryland, U.S.A.} \\
\footnotesize{$^4$Epidemiology Unit, National Institute of Allergy and Infectious Diseases, Bethesda, Maryland, U.S.A.} \\
\footnotesize{$^5$Critical Care Medicine Department, Clinical Center, National Institutes of Health, Bethesda, Maryland, U.S.A.} \\
\footnotesize{$^6$Division of Infectious Diseases, Emory University School of Medicine, Atlanta, Georgia, U.S.A.} \\
\footnotesize{$^{7}$Susan and Henry Samueli College of Health Sciences, University of California, Irvine, California, U.S.A.}\\
\footnotesize{$^{8}$Orange County Health Care Agency, Santa Ana, California, U.S.A.}\\
\footnotesize{$^{9}$KCS Health Center, Buena Park, California, U.S.A.}\\
\footnotesize{$^{10}$Center for Complex Biological Systems, University of California, Irvine, California, U.S.A.}\\
\footnotesize{$^{\dagger}$Corresponding author: {\tt vminin@uci.edu}}

\end{center}

\label{firstpage}

\begin{abstract}
Mechanistic models fit to streaming surveillance data are critical to understanding the transmission dynamics of an outbreak as it unfolds in real-time.
However, transmission model parameter estimation can be imprecise, and sometimes even impossible, because surveillance data are noisy and not informative about all aspects of the mechanistic model.
To partially overcome this obstacle, Bayesian models have been proposed to integrate multiple surveillance data streams. 
We devised a modeling framework for integrating SARS-CoV-2 diagnostics test and mortality time series data, as well as seroprevalence data from cross-sectional studies, and tested the importance of individual data streams for both inference and forecasting.
Importantly, our model for incidence data accounts for changes in the total number of tests performed.
We model the transmission rate, infection-to-fatality ratio, and a parameter controlling a functional relationship between the true case incidence and the fraction of positive tests as time-varying quantities and estimate changes of these parameters nonparametrically.
We compare our base model against modified versions which do not use diagnostics test counts or seroprevalence data to demonstrate the utility of including these often unused data streams.
We apply our Bayesian data integration method to COVID-19 surveillance data collected in Orange County, California between March 2020 and February 2021 and find that 32--72\% of the Orange County residents experienced SARS-CoV-2 infection by mid-January, 2021.
Despite this high number of infections, our results suggest that the abrupt end of the winter surge in January 2021 was due to both behavioral changes and a high level of accumulated natural immunity.
\end{abstract}

\clearpage

\section{Introduction}
\doublespacing
\label{sec:intro}
SARS-CoV-2 is a human coronavirus associated with high morbidity and mortality that caused a pandemic in 2020 \citep{Cummings2020, Wu2020CDC, Song2020}.
Like other human coronaviruses, SARS-CoV-2 is transmitted person to person through close contact and has high transmission potential in crowded indoor settings and around activities that generate aerosols \citep{whocoronavirustrans}.
In the early stages of the COVID-19 pandemic, transmission dynamics modeling played an important role in alerting the public about the potential dangers of unmitigated virus spread \citep{prem2020effect, ferguson2020report9, davies2020}.
At later pandemic stages, this modeling helped evaluate intervention effectiveness \citep{Knockeabg2021} and to quantify transmission advantages of genetic variants \citep{Davieseabg2021}.
We develop models that integrate diagnostics test and mortality time series data with cross-sectional seroprevalence data to estimate underlying transmission dynamics and forecast future case and death counts.
These models are flexible in the data sources they incorporate.
By comparing the forecasting capabilities of these models, we aim to investigate which data streams should be used in future modeling efforts.
\par
Differences in mitigation strategies, surveillance efforts, and population characteristics across countries and even across different regions within one country prompted development of regional modeling of SARS-CoV-2 transmission \citep{anderson2020, miller2020, morozova2021one, irons2021estimating}.
However, neither national nor subnational/regional modelers fully integrate all surveillance data available to them, because inclusion of each additional data source leads to an increase in model complexity, which complicates statistical inference and reduces computational efficiency of this inference.
In addition, including more data sources necessitates additional modeling assumptions and risks specification in one part of the model influencing inference in other aspects of the model.
Modelers were faced with many questions about which data to use, which data to ignore, and how to best integrate them into their models.
Incorporating case incidence data into inference proved particularly problematic because a data generating model for cases needs to account for preferential sampling of symptomatic individuals and dependence of case counts on the number of diagnostic tests performed, which significantly varies temporally and spatially.
However, even with delayed reporting, positive diagnostic tests (cases) are among the earliest indicators of changing disease dynamics, so we hypothesize that taking advantage of this source of information could be important for producing timely forecasts and for policy decision-making.
Similarly, properly incorporating seroprevalence data into models may improve the accuracy of a model's estimations of the underlying cumulative number of infections, which, in part, drives the effective reproduction number and is crucial for forecasting.
We investigate these ideas by fitting and comparing the forecasting abilities of multiple models, some of which use these data, while others do not.
\par
We show how to fit a mechanistic model of SARS-CoV-2 spread to incidence and mortality time series, while accounting for the time-varying number of diagnostic tests performed.
The mechanistic model is a standard ordinary differential equation (ODE) model that describes changes in the proportions of the population residing in model compartments.
Death counts are modeled with a negative binomial distribution that allows for over-dispersion often observed in surveillance data.
Our first innovation is the model for cases, where we use a flexible beta-binomial distribution, whose mean is a product of the total number of tests performed and a non-linear function of unobserved infections modeled by the ODE model.
This ensures that our estimates are not unduly influenced by large fluctuations of COVID-19 diagnostic test positivity fractions.
Our second innovation is nonparametric estimation of time-varying parameters that control both the transmission model and surveillance model.
By allowing our model to adapt to temporal changes in transmission and surveillance, we can identify how changes in policy affected the near-term progression of the outbreak.
Our third contribution is a careful assessment of the usefulness of various data streams in the context of a case study in Orange County, California, USA.
\par
To benchmark the ability of our model to capture temporal trends in transmission dynamics, we use simulated data to compare our estimation of changes in the effective reproductive number to analogous estimates produced by \texttt{epidemia}, a simpler semi-parametric method \citep{epidemia}.
This comparison shows that failing to account for fluctuations in the number of diagnostic tests performed can result in misleading inferences about effective reproduction numbers, a critical quantity in infectious disease epidemiology.
Our data integration approach allows our model to provide a much more detailed picture of the spread of an infectious disease beyond the effective reproduction number, including estimating the time-varying infection fatality ratio, the total number of infected individuals, and changes in testing patterns.
Data integration further allows us to produce reasonable short-term forecasts of deaths and testing positivity.
We demonstrate these enhanced capabilities
by fitting our compartmental model to COVID-19 surveillance data collected in Orange County, California --- the sixth most populous county in the United States of America (U.S.A.), with an estimated 3.2 million inhabitants as of 2019 \citep{orangecensus}.
We analyze Orange County surveillance data collected between March 30, 2020 and February 14, 2021, prior to the start of widespread vaccine availability.
We find both basic and effective reproductive numbers varied widely during the first year of the pandemic, which is expected in light of implementation and subsequent relaxation of mitigation measures.
We compare several modifications to our primary model which omit negative diagnostic tests or seroprevalence data.
We demonstrate that our models produce reasonable short term (up to 4 weeks ahead) probabilistic forecasts of mortality, but that different data streams may be more or less useful during different periods of the pandemic.

\section{Methods}
\label{sec:methods}

\subsection{Data}
\label{subsec:data}
\begin{figure}
\includegraphics[width=1.0\textwidth]{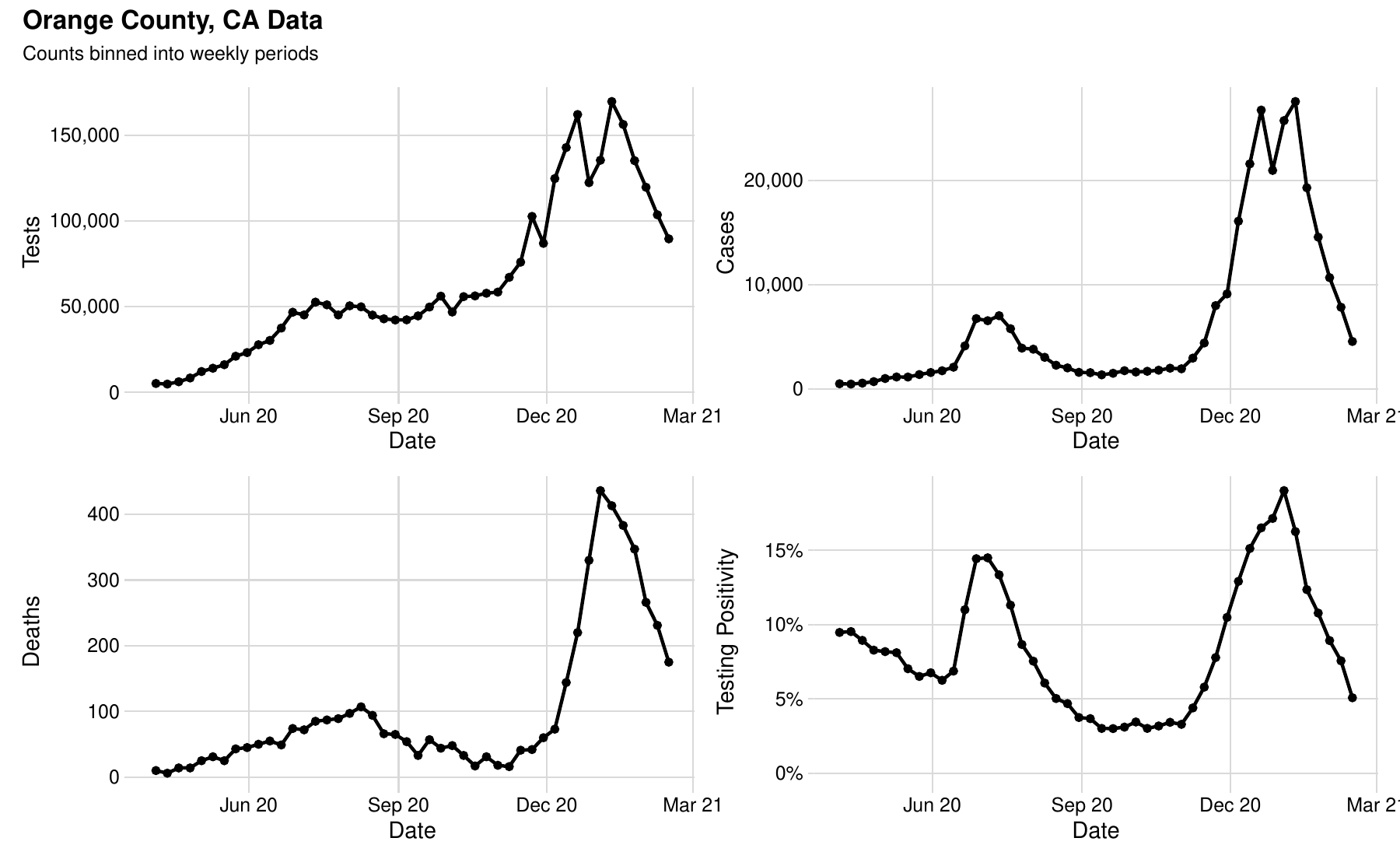}
\caption{
COVID-19 surveillance data from Orange County, CA.
The figure shows weekly counts of tests, cases (positive tests), reported deaths due to COVID-19, as well as testing positivity.}
\label{oc:data}
\end{figure}

We start with time series of daily numbers of SARS-CoV-2 diagnostic tests (positive and negative), case counts (positive tests), and deaths observed over some time period of interest.
We aggregate the three types of counts in weekly intervals.
Figure~\ref{oc:data} shows such a collection of aggregated time series for Orange County, CA, corresponding to the observation period spanning days between March 30, 2020 and February 14, 2021.
We end our modeling period in February because vaccines became more widely available around this time, and our model does not account for vaccine-induced immunity.
The data was compiled from anonymized individual test results provided by the Orange County Health Care Agency (OCHCA).
We define cases as either confirmed or presumed COVID-19 diagnoses that have been officially reported to the OCHCA.
We used specimen collection dates and dates of deaths to tabulate test, case, and death counts.
We denote the vector of binned tests by $\mathbf{T} = (T_1, \dots, T_L)$, the vector of case counts by
$\mathbf{Y} = (Y_1, \dots, Y_L)$, and the vector of deaths by $\mathbf{M} = (M_1, \dots, M_L)$, where the weeks are indexed by $l$ and $L = 48$ is the total number of weeks.
Additionally, we use data from \cite{Bruckner2021}, a study conducted to estimate the seroprevalence in Orange County from a population-representative sample, which consists of 343 seropositive cases ($Z_{l^*}$) among 2979 ($U_{l^*}$) tests conducted between July 10 and August 16, 2020.
For simplicity, we lump all seroprevalence test dates to a single time point --- August 16, 2020 --- corresponding to week $l^*=20$.
To formulate the surveillance model for cases $\mathbf{Y}$, deaths $\mathbf{M}$, and seropositive cases $Z_{l^*}$, we first need a model for latent trajectories of incidence and prevalence of SARS-CoV-2 infections.

\subsection{Transmission model}
\label{subsec:model}
To model latent incidence and prevalence trajectories, we divide all individuals in the population of Orange County, CA into five compartments: $S$ = susceptible individuals, $E$ = infected, but not yet infectious individuals, $I$ = infectious individuals, $R$ = recovered individuals, $D$ = individuals who died due to COVID-19.
Possible progressions of an individual through the above compartments are depicted in Figure~\ref{fig:seiird}.
We model the time-evolution of the proportions of individuals occupying the above compartments with a set of deterministic ordinary differential equations (ODEs).
For simplicity, we assume a homogeneously mixing population of fixed size \( N \), although it is possible to relax these assumptions, and we also assume that recovery confers immunity to subsequent infection over the duration of the modeling period.
Let $\bX(t) = \left(S(t), E(t), I(t), R(t), D(t) \right)^T$ denote the population proportion in each compartment at time $t$, and let $\bX(t_0) = \bx_0 $ denote the population proportions at time $ t_0 $, the start of the modeling period.
By convention, we model the population at risk, i.e., those individuals who may still move throughout the model compartments.
Hence, we take $R(t_0) = D(t_0) = 0$ and normalize $ \bX(t_0) $ so that $ \bX(t_0)^T\boldsymbol{1} = 1$, where $\boldsymbol{1} = (1,1,1,1,1)^T$.
Since we want to fit this model to incidence data, it is convenient to also keep track of the cumulative proportion of the population that experiences transitions between compartments from $t_0$ to $t$: $ \bN(t) = \left(N_{SE}(t), N_{E I}(t), N_{I R}(t), N_{I D}(t)\right)^T$.
To describe mathematically how vectors $\bX(t)$ and $\bN(t)$ change through time, we first define rates of transitions between compartments, with possible transitions corresponding to the arrows in Figure~\ref{fig:seiird}:

\begin{equation}
\begin{aligned}
\lambda_{SE}(S, I, t) &= \frac{\beta(t)}{N} S I,\\
\lambda_{I R}(I, t) &= \left(1-\eta(t)\right) \nu I,
\end{aligned}
\qquad \quad \quad
\begin{aligned}
\lambda_{E I}(E) &= \gamma E,\\
\lambda_{I D}(I, t) &= \eta(t) \nu I,
\end{aligned}
\end{equation}
where $\beta(t)$ is the transmission rate, which varies over time, \( N \) is the constant population size, $1/\gamma$ is the mean latent period duration, $1/\nu$ is the mean infectious period duration, and $\eta(t)$ is the infection-to-fatality ratio (IFR), which varies over time.
The time-varying transmission rate allows our model to capture the effects of interventions and changes in human behavior, while the time-varying IFR should capture changes in age profiles of infected individuals and stress on healthcare providers during surges.
We demonstrate this property in Supplementary Section~\ref{sec:hetero}.
\par
Equipped with the population-level transition rates, we define ODEs for our model:
\begin{equation}
\label{eqn:model_ODEs}
\begin{aligned}
\deriv{S}{t} &= -\lambda_{SE}(S, I, t),\\
\deriv{E}{t} &= \lambda_{SE}(S, I, t)  - \lambda_{E I}(E),\\
\deriv{I}{t} &=  \lambda_{E I}(E) - \lambda_{I R}(I, t) -  \lambda_{I D}(I, t),\\
\deriv{R}{t} &=   \lambda_{I R}(I, t),\\
\deriv{D}{t} &= \lambda_{I D}(I, t),
\end{aligned}
\qquad \qquad \qquad \qquad
\begin{aligned}
\deriv{N_{SE}}{t} &= \lambda_{SE}(S, I, t),\\
\deriv{N_{E I}}{t} &=  \lambda_{E I}(E),\\
\deriv{N_{I R}}{t} &= \lambda_{I R}(I, t),\\
\deriv{N_{I D}}{t} &= \lambda_{I D}(I, t),\\
\
\end{aligned}
\end{equation}
subject to initial conditions $ \bX(t_0) = \bx_0 $ and $ \bN(t_0) = \mathbf{0}$,
where $\bx_0 = (S_0, E_0, I_0, R_0, D_0)$ are initial compartment proportions.
We set $R_0 = 0$ and $D_0 = 0$, because these proportions do not play a role in future dynamics of the epidemic, leaving $S_0$, $E_0$, and $I_{0}$ as free model parameters.

The above equations are redundant, and typically only the prevalence ODEs in the left column are used in mathematical modeling.
However, the cumulative incidence/transition representation of the model, shown by the ODEs in the right column, is useful for statistical modeling of infectious disease dynamics \citep{breto2011compound}.
In practice, we solve the subset of the above ODEs that are needed to track $\bX(t)$ and the parts of $\bN(t)$ that ``connect'' our transmission model to data.
We proceed to make this connection in the next subsection.

\begin{figure}[htbp]
	\centering
	\resizebox {0.7\linewidth} {!} {
		\begin{tikzpicture}
		\draw (0, -1.25) circle(0.5) node (S) {$ S$};
		\draw (2, -1.25) circle(0.5) node (E) {$ E $};
		\draw (4, -1.25) circle(0.5) node (I) {$ I $};
		\draw (6, -2) circle(0.5) node (D) {$ D $};
		\draw (6, -0.5) circle(0.5) node (R) {$ R $};

		\draw [thick,->,shorten >=3mm,shorten <=3mm] (S) -- (E) node[midway,above] {};
		\draw [thick,->, shorten >=3.5mm, shorten <=3mm] (E) -- (I) node[midway,above] {};
		\draw [thick,->,shorten >=3mm, shorten <= 3mm] (I) -- (R) node[midway,above] {};
		\draw [thick,->,shorten >=3mm, shorten <= 3mm] (I) -- (D) node[midway,above] {};
		\end{tikzpicture}
	}
	\caption{Model diagram depicting possible progressions between infection states. The model compartments are as follows: susceptible $ (S) $, infected, but not yet infectious ($ E $), infectious ($I$), recovered ($ R $), and deceased ($ D $). }
	\label{fig:seiird}
\end{figure}

\subsection{Surveillance model}
We fit our transmission model to seroprevalence data and two time series: numbers of new cases and deaths reported during some pre-specified time periods (e.g., weeks).
We do not model changes in the numbers of diagnostic tests performed.
Rather, we condition on test counts in the specification of the sampling model for the vector of case counts, which describes the probability of the observed case count given the observed number of tests and unobserved/latent incidence of cases over each time interval.
First, we assume that, conditional on $\bN(t)$, case and death counts are independent of each other and across time intervals, because they are just noisy realizations of information encoded by $\bN(t)$.
This leaves us with formulating models for cases and deaths in each individual observation interval.
\par
Consider the number of deaths $M_l$ observed in time interval $(t_{l-1}, t_l]$, where $l = 1,\dots, L$.
Since our ODEs track the latent cumulative fraction of deaths $N_{I D}(t_l)$, we can compute
$\Delta N_{I D}(t_l) = N_{I D}(t_l) - N_{I D}(t_{l-1})$ --- the latent fraction of the population that died in the interval $(t_{l-1}, t_l]$.
We model the observed death count $M_l$ as a realization from the following negative binomial distribution:
\begin{align}
\label{eqn:mortalityemission}
M_l \sim \text{Negative binomial}\left (\mu^D_l =  \rho_D \times N \times \Delta N_{I D}(t_l),\ {\sigma^2_l}^D = \mu^D_l (1 + \mu^D_l / \phi_D )\right ),
\end{align}
where $N$ is the population size, $ \mu^D_l $ and $ {\sigma^2_l}^D $ are the mean and variance of the negative binomial distribution, $ \rho_D \in [0,1]$ is the mean overall death detection probability, and $ \phi_D > 0 $ is an over-dispersion parameter.
Informally, our mortality model says that, on average, the observed number of deaths, $ M_l $, is a fraction of the true death count estimated by the model, $N \times \Delta N_{I D}(t_l)$,  with some noise due to underreporting, delayed reporting, and sampling variability.
\par
Next, we develop a model for the number of positive tests (cases), $Y_l$, observed in the time interval $(t_{l-1}, t_l]$.
We start with a simple binomial model with per-test positivity probability $\psi_l$:
\[
Y_l \mid \psi_l \sim \text{Binomial}(T_l, \psi_l),
\]
where $T_l$ is the number of COVID-19 diagnostic tests administered during the time interval $(t_{l-1}, t_l]$.
We use another layer of randomness to account for unobserved factors affecting positivity probabilities (e.g., variable testing guidelines and test shortages) and assume that the positivity probability in interval $(t_{l-1}, t_l]$ follows the beta distribution:
\begin{equation}
\psi_{l}  \sim \text{Beta}\left (\phi_C \mu^C_l,\ \phi_C \left (1 - \mu^C_l \right )\right ),
\label{eqn:positivity}
\end{equation}
where $ \phi_C $ is an over-dispersion parameter and $\mu^C_l$ is the mean test positivity probability.
We assume that mean test positivity odds is proportional to the unobserved odds of transitioning from exposed to infectious, $\Delta N_{E I}(t_l) = N_{E I}(t_l) - N_{E I}(t_{l-1})$ in interval $(t_{l-1}, t_l]$:

\begin{equation}
\frac{\mu^C_l}{1- \mu^C_l}= e^{\alpha_l} \cdot \left( \frac{\Delta N_{EI}(t_l)}{1 - \Delta N_{EI}(t_l)}\right),
\label{eqn:testlogoddsmean}
\end{equation}
where $\alpha_l > 0$.
This functional form ensures that, on average, the probability of detecting a SARS-CoV-2 infection grows with the population incidence.
Parameter $\alpha_l$ can be thought of as an effect of testing guidelines and practices.
A model with $\alpha_l \approx 0$ (i.e., $\mu^C_l \approx \Delta N_{\Ie\Is}(t_l)$) says that in interval $(t_{l-1}, t_l]$ testing is done approximately by sampling individuals uniformly at random, so that the positivity probability over a time interval $ l $ is equal to the fraction of the population that transitions from the latent to infectious state.
As we increase $\alpha_l$ above 0, the model mimics preferential testing of individuals who are more likely to have severe infection (e.g., testing only individuals with certain symptoms).

\par We can streamline our surveillance model for case counts by integrating over positivity probabilities and arriving at the following beta-binomial distribution:
\begin{align}
\label{eqn:caseemission}
Y_l\mid \mu_l^C,\phi_C \sim \text{Beta-binomial}\left (T_l,\ \phi_C \mu^C_l,\ \phi_C \left (1 - \mu^C_l\right )\right ).
\end{align}
Properties of the beta-binomial distribution imply that $\text{E}(Y_l) = T_l \times \mu^C_l$.
This means that our model predicts that, on average, cases grow linearly with the number of diagnostic tests administered.
Keeping in mind our assumed relationship between $\mu^C_l$ and $\Delta N_{E I}(t_l)$, the average number of cases also grows with the accumulation of new infections.
Furthermore, the variance of the fraction of tests that are positive under the beta-binomial distribution is
\[
\Var(Y_l/T_l\mid T_l, \mu_l^C, \phi_C) = \frac{\mu_l^C(1-\mu_l^C)}{T_l}\left (1 + \frac{T_l - 1}{\phi_C+1}\right ),
\]
where the variance under an analogous pure binomial model would be $ \mu_l^C(1-\mu_l^C)/T_l $. Hence, the over-dispersion parameter, $ \phi_C $, can be interpreted in terms of the excess variance of the beta-binomial model relative to a pure binomial distribution.
In summary, our beta-binomial distribution for observed case counts ensures that we do not confuse increase in testing for increase in SARS-CoV-2 incidence and implicitly allows for heterogeneity in the mean test positivity probability.

Finally, we model the number of observed seropositive cases $Z_{l^*}$ among $U_{l^*}$ tests with a binomial distribution:
\begin{equation}
\label{eqn:seroprevemission}
    Z_{l^*} \sim \text{Binomial}\left( U_{l^*}, \frac{R_{l^*}}{S_{l^*} + E_{l^*} + I_{l^*} + R_{l^*}} \right).
\end{equation}

This simple model assumes the seroprevalence data comes from a high-quality study based on random sampling, which does not exhibit the problems observed in the testing data.

In addition, we also consider a more typical approach to modeling observed cases, which is not conditional on tests and is similar to \eqref{eqn:mortalityemission}.

\begin{equation}
    \label{eqn:altcaseemission}
    Y_l \sim \text{Negative binomial}\left (\mu^Y_l =  \rho_l^Y \times N \times \Delta N_{E I}(t_l),\ {\sigma^2_l}^Y = \mu^Y_l (1 + \mu^Y_l / \phi_Y )\right ),
\end{equation}

where $N$ is the population size, $ \mu^Y_l $ and $ {\sigma^2_l}^Y $ are the mean and variance of the negative binomial distribution, $ \rho_l^Y \in [0,1]$ is the mean case detection probability, which varies over time, and $ \phi_Y > 0 $ is an over-dispersion parameter.

\subsection{Putting all the pieces into a Bayesian model}
\label{subsec:limitations}
We now describe our inferential Bayesian procedure.
First, we re-parameterize our model by replacing $\beta(t)$ with a basic reproductive number $R_0(t) = \beta(t) / \nu$.
We parameterize each of our time-varying parameters, $R_0(t)$, $\eta(t)$, $\alpha(t)$, $\rho_l^Y(t)$ as piecewise constant functions, where each vector defining the constants \textit{a priori} follows a Gaussian Markov random field (GMRF).

More precisely, we define the auxiliary vectors:
\[
\brtilde = (\tilde{R}_{0,1}, \tilde{R}_{0,2}, \ldots, \tilde{R}_{0,L}) \text{, } \betatilde = (\etatilde_1, \etatilde_2, \ldots, \etatilde_L) \text{, } \balphatilde = (\alphatilde_1, \alphatilde_2, \ldots, \alphatilde_L)  \text{, and } \brhoytilde = (\brhoytilde_1, \brhoytilde_2, \ldots, \brhoytilde_L),
\]
which follow the Gaussian Markov random field priors:

\begin{equation}
\begin{aligned}
	\tilde{R}_{0,l} \sim& N\left(\tilde{R}_{0, l - 1}, \sigma_{R_0}^2\right)\text{, where } l=2,\ldots,L \text{ and } \tilde{R}_{0,1} \sim N\left(\mu_{R_{01}},\sigma^2_{R_{01}}\right),\\
	\etatilde_{l} \sim& N\left(\etatilde_{l - 1}, \sigma_\eta^2\right)\text{, where } l=2,\ldots,L \text{ and } \etatilde_1 \sim N\left(\mu_{\eta_1}, \sigma^2_{\eta_1}\right),\\
	\alphatilde_{l} \sim& N\left(\alphatilde_{l - 1}, \sigma_\alpha^2\right)\text{, where } l=2,\ldots,L \text{ and } \alphatilde_1 \sim N\left(\mu_{\alpha_1},\sigma^2_{\alpha_1}\right),\\
	\rhoytilde_{l} \sim& N\left(\rhoytilde_{l - 1}, \sigma_{\rho^Y}^2\right)\text{, where } l=2,\ldots,L \text{ and } \rhoytilde_1 \sim N\left(\mu_{\rho^Y_1}, \sigma^2_{\rho^Y_1}\right),
 \label{eqn:GMRF}
\end{aligned}
\end{equation}
and define the piecewise constant functions:
\begin{align}
	R_0(t) =& \sum_{l=1}^L \exp{\left( \tilde{R}_{0,l} \right) \1{t \in (t_{l-1}, t_l]}},\\
	\eta(t) =& \sum_{l=1}^L \frac{\exp{ \left(\etatilde_l\right) \1{t \in (t_{l-1}, t_l]} }}{\exp{\left( \etatilde_l \right) \1{t \in (t_{l-1}, t_l]}} + 1},\\
	\alpha(t) =& \sum_{l=1}^L \exp{\left( \alphatilde_l \right)\ 1{t \in (t_{l-1}, t_l]} },\\
	\rho^Y(t) =& \sum_{l=1}^L \frac{\exp{\left( \rhoytilde_l \right) \1{t \in (t_{l-1}, t_l]} }}{\exp{\left( \rhoytilde_l \right) \1{t \in (t_{l-1}, t_l]} } + 1}.
\end{align}

In addition, we parameterize initial compartment fractions as $ S_0 $, $ I_0 = \til{I}_0(1 - S_0) $, $ E_0 = (1 - \til{I}_0)(1 - S_0) $, $R_0 = 0$, $D_0 = 0$, where $\til{I}_0 = I_0 / (E_0 + I_0)$.
This construction allows us to specify independent prior distributions for $S_0$ and $\til{I}_0$ while preserving the sum-to-one constraint on the original initial compartmental fractions.
Next, we collect all our model parameters into a vector $\boldsymbol{\theta} = (S_0,  \tilde{I}_{0}, \brtilde, \gamma, \nu, \betatilde, \rho^D, \phi_D, \balphatilde, \phi_C, \sigma_{R_0}, \sigma_\eta, \sigma_\alpha)$.
When using the traditional case-emission model \eqref{eqn:altcaseemission}, $ \brhoytilde $,  $ \sigma_{\rho^Y} $, and $\phi_Y $ are substituted for $ \balphatilde $, $ \phi_C $, and $ \sigma_\alpha $.
Our probabilistic construction described above implies that the likelihood function --- probability of observing incidence, mortality, and seroprevalence data --- can be written in the following way:

\begin{equation*}
\text{Pr}(\bM, \bY,  Z_{l^*} \mid \btheta ) = \text{Pr}(\bM \mid \btheta ) \text{Pr}(\bY \mid \btheta ) \text{Pr}(Z_{l^*} \mid \btheta ) =   \prod_{l=1}^{L}\Pr(M_l\mid\btheta) \Pr(Y_l\mid\btheta) \Pr(Z_{l^*}\mid\btheta),
\end{equation*}
where $\text{Pr}(M_l \mid \btheta)$, $\text{Pr}(Y_l \mid \btheta)$, and $\text{Pr}(Z_{l^*} \mid \btheta)$ are the probability mass functions given by \eqref{eqn:mortalityemission}, \eqref{eqn:caseemission} or \eqref{eqn:altcaseemission}, and \eqref{eqn:seroprevemission} respectively.
\par
We encode available information about our model parameters in a prior distribution with density $\pi(\btheta)$.
We assume that all univariate non-GMRF distributed parameters are \textit{a priori} independent and list our prior assumptions in Table~\ref{table:priors}.
Since our model is highly parametric, we rely on informative prior distributions that we parameterize using existing scientific studies.
We base all our inferences and predictions on the posterior distribution of all model parameters:
\begin{equation}
\pi(\btheta\mid \bM,\bY, \bZ) \propto \text{Pr}(\bM, \bY, \bZ \mid \btheta ) \pi(\btheta).
\end{equation}

We sample from this posterior using the No-U-Turn Sampler \citep{NUTS} as implemented in the Turing Julia package \citep{turing}.
Model code and data are available at the following GitHub repository: \url{https://github.com/damonbayer/semi_parametric_COVID_19_OC_model}.

\section{Results}
\label{sec:results}
\subsection{Simulation Study}
To validate our model, we simulated 200 datasets with parameters given in Supplementary Table~\ref{table:simulation_parameters}.
An example of one of these datasets is presented in Figure~\ref{fig:simulated_binned_data_plot}.
The number of tests at each time point is the same as in the Orange County data set, and the parameters were deliberately chosen to produce data similar to the Orange County data.
Priors used for these model fits are the same as those used in the Orange County model (see Table~\ref{table:priors}).
The prior and posterior distribution of the model fit to the single simulated dataset from Figure~\ref{fig:simulated_binned_data_plot} are presented in Supplementary Figures~\ref{fig:single_generated_quantities_simulation_scalar_plot}--\ref{fig:single_generated_quantities_simulation_time_varying_plot}.
For each simulated dataset, we used four Markov chains run in parallel to draw a total of 1000 posterior samples.
In this single dataset example, most of the scalar parameter posteriors shift slightly toward the true parameters compared to the priors, without much posterior variance contraction relative to the prior.
We define posterior variance contraction as one minus the ratio of standard deviation of the posterior and the prior, where negative contraction indicates that the posterior is wider than the prior, and 100\% contraction indicates that the posterior is a degenerate distribution.
For the time-varying parameters and compartments, the variance contraction and shift are much more apparent.
We further explore our simulation study results with summary measurements, presented in Supplementary Figures~\ref{fig:generated_quantities_simulation_scalar_coverage_plot}--\ref{fig:generated_quantities_simulation_compartment_coverage_plot}.
Figure~\ref{fig:generated_quantities_simulation_scalar_coverage_plot} shows the coverage of the posterior 80\% credible intervals constructed for the scalar parameters in the 200 simulated datasets.
Most parameters demonstrate nearly 100\% coverage, except for \( \phi_C \), which shows approximately 90\% coverage, which is still above the nominal 80\%.
Similarly, Figure~\ref{fig:generated_quantities_simulation_time_varying_coverage_plot} displays coverage of the posterior 80\% credible intervals constructed for the time-varying parameters, with only one time point for \( \alpha \) demonstrating less than nominal coverage.
We observe slightly less conservative coverage when examining the latent compartments in Figure~\ref{fig:generated_quantities_simulation_compartment_coverage_plot}, with the \( D \) compartment falling to around 60\% coverage at some points.
In addition to coverage, we also consider posterior contraction.
Figure~\ref{fig:generated_quantities_simulation_scalar_shrinkage_plot} shows contraction of the scalar parameters, with most parameters demonstrating nearly no contraction.
Notable exceptions to this are \( \sigma_{R_0} \), and \( \sigma_\alpha \), which exhibit positive contraction.
Figure~\ref{fig:generated_quantities_simulation_time_varying_shrinkage_plot} shows contraction of the time-varying parameters, with all parameters exhibiting a high amount of positive contraction.
Similarly, the latent compartments also demonstrate a large degree of positive contraction in Figure~\ref{fig:generated_quantities_simulation_compartment_shrinkage_plot}.

We used the \texttt{epidemia} R package \citep{epidemia} to fit a state-of-the-art method for effective reproduction number estimation (\( R_t \)), to the same 200 simulated datasets.
This semi-mechanistic method does not attempt to estimate the unobserved number of susceptible and recovered individuals and does not account for changes in testing volume.
See Supplementary Section \ref{epidemia_section} for a brief description of the \texttt{epidemia} statistical model.
Metrics based on estimates of \( R_t \) from the true model and \texttt{epidemia} for the simulation study are presented in Figure~\ref{fig:rt_comparison_metrics_plot}.
We assess the envelope, which is the proportion of time points which the 80\% posterior credible interval contains the true \( R_t \) value specified in the simulation.
Mean credible interval width (MCIW) is calculated as the mean of credible interval widths across time points within a simulation replication.
Absolute deviation is a measure of bias, and is calculated as the mean of the absolute difference between the posterior median and the true \( R_t \) value at each time point.
The mean absolute sequential variation (MASV) is calculated as the mean of the absolute difference between the posterior median at a time point and the posterior median at the previous time point.
In all metrics, the true model fit achieves the superior result.
The envelope for \texttt{epidemia} is typically around 50\%, while it is near 100\% for the true model.
The mean credible interval width for \texttt{epidemia} (around 0.50) is larger than for the true model (around 0.35).
The \texttt{epidemia} results also indicate more bias compared to the true model, as measured by the absolute deviation (0.20 for epidemia vs 0.05 for the true model).
Additionally, the true model has a mean absolute sequential variation close to that of the simulated parameters (around 0.085), while the MASV reported by \texttt{epidemia} is larger (around 0.10).
\par
\subsection{Application to Orange County, California Data}
Next, we apply our Bayesian inferential procedure to COVID-19 surveillance data collected in Orange County, California between March 30, 2020 and January 17, 2021.
We again used four Markov chains run in parallel to draw a total of 1000 posterior samples.
By the end of the modeling period, approximately 4\% of Orange County residents were at least partially vaccinated.
% https://web.archive.org/web/20210119193138/https://occovid19.ochealthinfo.com/vaccines-administered-oc
Because our model does not incorporate vaccination directly, it doesn't make sense to use our model beyond January 2021.
Throughout the modeling period, a variety of non-pharmaceutical interventions were enacted and sometimes lifted at the state, county, and city level.
Notably, in-person school closures, indoor dining bans, and mask mandates were in effect for most or all of this period.
Fitting the model took approximately 7.5 hours to generate 1,000 posterior samples from 4 chains, totaling around 29 CPU hours.
These run times show that our model can be fit frequently enough to be useful for a real-time policy response.
This number of chains and posterior samples resulted in a satisfactory effective sample size for posterior inference.
Convergence and mixing were assessed using potential scale reduction factors, effective posterior sample sizes, and traceplots of model parameters, which are presented in Appendix \ref{sec:convergence-diagnostics}.
Our main interest is in understanding differences in transmission dynamics and surveillance efforts throughout this period.
Since our model is highly parametric, we used existing knowledge of SARS-CoV-2 transmission dynamics to formulate informative priors for all model parameters that we list in Table~\ref{table:priors}.
We briefly highlight some of our assumptions.
Our priors for the initial compartment sizes reflect our belief that the number of infections was small, but potentially underreported by a factor of 10,  at the beginning of the pandemic.
Since our observation period starts close to the stay-at-home order taking effect in California, we assume that the March 2020 basic reproduction number should be around 1.0 to reflect reduced contacts during this period.
Lengths of latent and infectious periods \textit{a priori} assumed to be 0.8 and 1.2 weeks respectively, with substantial variance.
Based on Orange County, CA seroprevalence study, we assume initial infection-to-fatality ratio to be around 0.4\% \citep{Bruckner2021}.
We compare reported deaths in Orange County,  CA with estimates of U.S. county-specific excess mortality to set the prior for death reporting probability to be around 0.9 \citep{stokes2020assessing}.
Prior and posterior distributional summaries of all model parameters are available in Supplementary Figures~\ref{fig:scalar_sensitivity_plot} and \ref{fig:time_varying_sensitivity_plot}.
These figures also contain the results of our sensitivity analysis, where we examined the effects of our prior assumptions on our inference.
Our main conclusion is our results are not sensitive to reasonable prior perturbations.

\begin{figure}[htbp]
    \centering
    \includegraphics[width=1.0\textwidth]{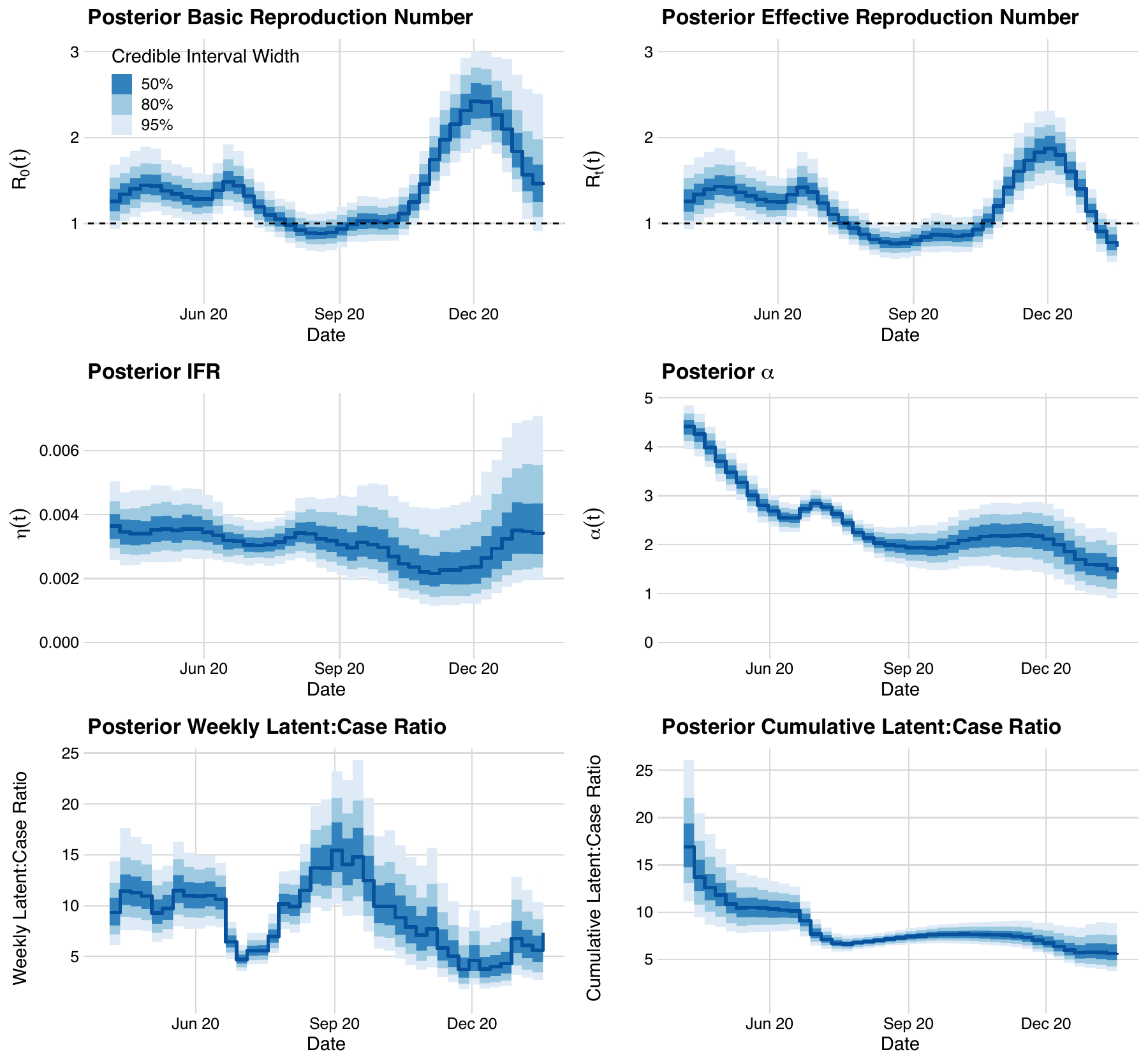}
    \caption{Posterior distributions of the time-varying basic reproductive number $R_0$, effective reproductive number $R_e$, infection-to-fatality ratio (IFR), proportion in the proportional log-odds model of the beta-binomial observational model for cases $\alpha$, weekly latent:case ratio, and cumulative latent:case ratio.
    Solid blue lines show point-wise posterior medians, while shaded areas denote 50\%, 80\%, and 95\% Bayesian credible intervals.}
    \label{fig:main_posterior_results_plot}
\end{figure}

\begin{figure}
    \centering
    \includegraphics{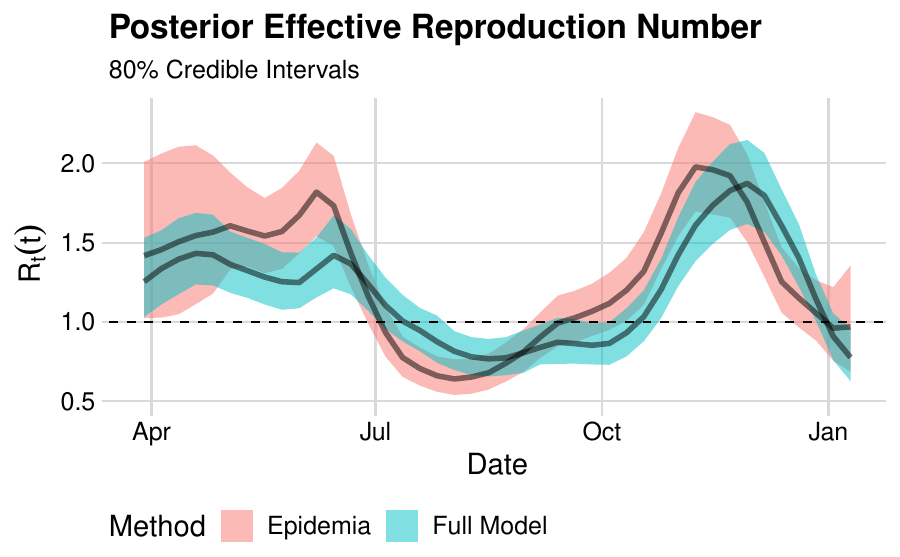}
    \caption{Posterior inference for the effective reproduction number from the full model and \texttt{epidemia} fit to the Orange County data.}
    \label{fig:rt_comparison_oc_data_plot}
\end{figure}

The upper-left plot of Figure~\ref{fig:main_posterior_results_plot} presents the posterior distribution of the basic reproductive number ($R_0$) for Orange County.
Throughout the late spring and summer, the basic reproductive number is estimated to be slightly above 1.0, with some probability of being below 1.0 in the early fall.
Beginning in October, the basic reproductive number begins to rise and surpasses 2.0 at the peak of the winter wave.
This rise in the fall may be associated with the school re-openings that occurred around this time. % https://www.ocregister.com/2020/09/15/heres-what-we-know-about-in-person-learning-plans-for-orange-county-public-school-districts/
Despite the high basic reproductive number throughout the modeling period, the upper-right plot of Figure~\ref{fig:main_posterior_results_plot} shows that the effective reproductive fell below 1.0 for much of the summer and again in January, following the winter surge, allowing us to separate the effects of reducing the average community contact rate and accumulated infection-induced immunity.
We also apply the \texttt{epidemia} method to the Orange County data and plot the results in Figure~\ref{fig:rt_comparison_oc_data_plot}.
From this, we observe that the two methods lead to similar conclusions about the posterior distribution of \( R_t \).
At all but one of the time points, the 80\% credible intervals of the posteriors from both methods overlap with one another, but the full model appears to generally produce smaller credible intervals, especially near the beginning of the fitting period.
Additionally, the full model produces a smoother posterior than the \texttt{epidemia} model.

\par
We proceed with describing inference results for the other two time-varying parameters in our model: infection-to-fatality ratio \( \eta \)
and the parameter $\alpha$ that governs the relationship between testing positivity and the true proportion of newly infected individuals in the population.
The posterior distribution of the infection-to-fatality ratio is presented in the middle-left plot of Figure~\ref{fig:main_posterior_results_plot}.
The IFR is estimated to be consistent over time, hovering around 0.3\%, but our estimates are less certain near the end of the modeling period.
This potential rise in IFR could have been caused by a combination of the overwhelmed healthcare system and the increasing prevalence of the Alpha variant at this time, which has been tied to more severe outcomes \citep{grint2021severity}.
The middle-right plot and bottom plots of Figure~\ref{fig:main_posterior_results_plot} present three perspectives on testing policy and case detection: the posterior \( \alpha \), the weekly latent:case ratio, and the cumulative latent:case ratio.
Generally, \( \alpha \), drifts lower over time, indicating that testing policy became less preferential toward selecting infected individuals as testing became more accessible.
This trend is reversed slightly during the summer and winter waves, which is reflected in the decreasing weekly latent:case ratio during these times.
The cumulative latent:case ratio also drifts lower over time, eventually arriving at a final cumulative latent:case ratio of 4:1 -- 9:1.
\par
\begin{figure}[htbp]
    \centering
    \includegraphics[width=1.0\textwidth]{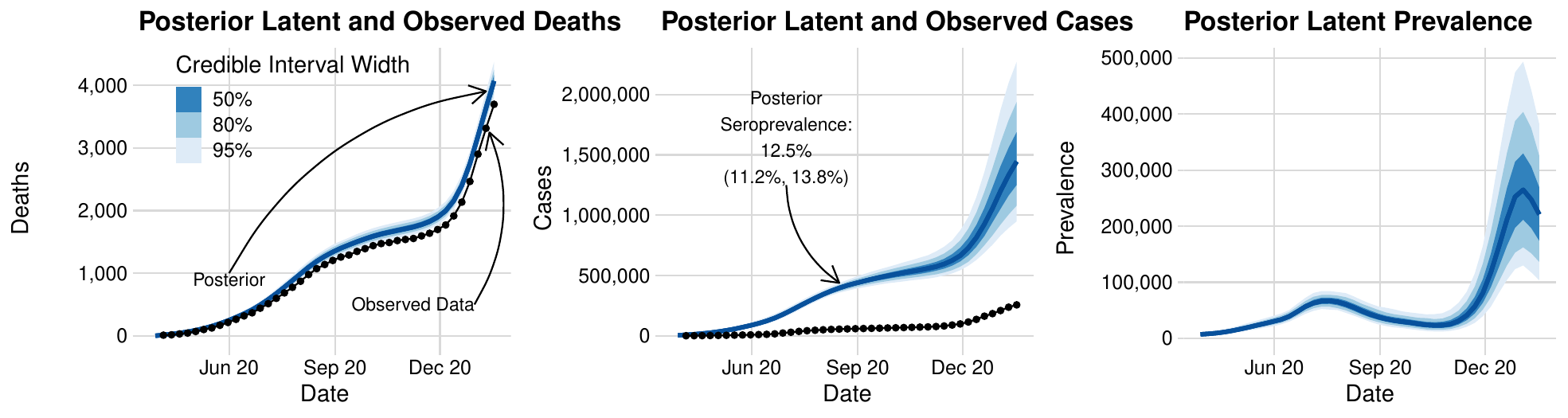}
    \caption{Latent and observed cumulative death (left) and incidence (center) trajectories and latent prevalence trajectories (right) in Orange County, CA (population 3.2 million).
    Solid blue lines show point-wise posterior medians, while shaded areas denote 50\%, 80\%, and 95\% Bayesian credible intervals.
    Black circles denote observed data.
    Note that the posterior predictive distributions are of latent deaths and cases are not forecasts of their observed counterparts.
    Forecasts are plotted in Figure~\ref{fig:combined_forecast_plot}.}
    \label{fig:dip_plot}
\end{figure}

We plot posterior medians and Bayesian credible intervals of the latent cumulative death counts ($N_{ID}(t)$) between March 2020 and January 2021, using three credibility levels shown in the left plot of Figure~\ref{fig:dip_plot}.
Reported death counts are shown as black circles in the same plot.
The plot reflects an overall death reporting rate of 87\% - 94\%.
The center plot of Figure~\ref{fig:dip_plot} shows the posterior distributions of the cumulative number of infections ($N_{SE}(t)$) occurred in Orange County, with the cumulative observed cases displayed as black circles.
We estimate that 32--72\% of Orange County residents experienced SARS-CoV-2 infection by mid-January 2021.
As in Figure~\ref{fig:main_posterior_results_plot}, this shows a cumulative latent:case ratio of 4:1 -- 9:1, with 1/3 -- 2/3 of all Orange County residents having been infected by the end of January 2021.
From this plot, we also note that our posterior estimate of seroprevalence in mid-August of 2020 (11.2\%-13.7\%) closely matches the 11.5\% estimate from \citep{Bruckner2021}. We explicitly used this seroprevalence data in our inference, so this is unsurprising.
The right plot of Figure~\ref{fig:dip_plot} shows the prevalence of SARS-CoV-2 infected individuals at a particular time, ($E(t) + I(t)$).
At the peak of the winter wave, we estimate that 7.8\% -- 14.9\% of Orange County residents had an active infection at the same time.
\par

\begin{figure}[htbp]
    \centering
    \includegraphics[width=1.0\textwidth]{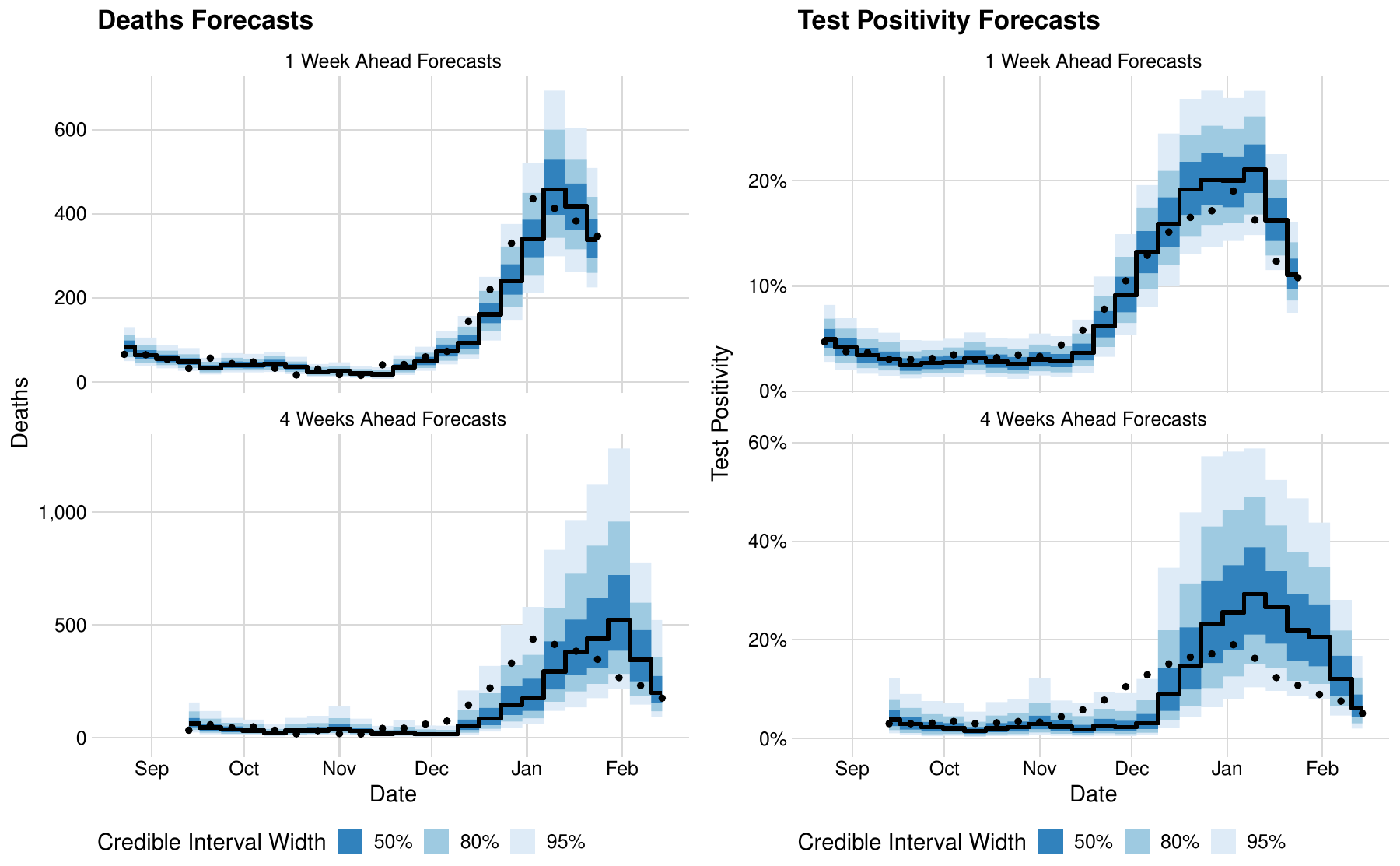}
    \caption{Forecast distributions for observed deaths (left column) and testing positivity (right column).
    Solid blue lines show point-wise posterior medians, while shaded areas denote 50\%, 80\%, and 95\% Bayesian credible intervals.
    Observed values are presented as black circles.}
    \label{fig:combined_forecast_plot}
\end{figure}

Finally, we turn to model-based forecasting of observable quantities.
In Figure~\ref{fig:combined_forecast_plot}, we present one week and four week ahead forecasts of observed deaths and test positivity.
The credible intervals shown for a given date are generated from a model that is fit to data from March 30, 2020 up to 1 week or 4 weeks prior to the given date.
The forecasts are produced by augmenting the posterior time-varying parameters by carrying forward the previous mean values in \eqref{eqn:GMRF} and solving the ODEs from \eqref{eqn:model_ODEs} into the future.
Since this model makes use of the seroprevalence data, we only produce forecasts for times after this data is available, beginning in late August 2020.
Because forecasting cases with this model is impossible without knowing how many tests will be conducted in the future, we focus on the positivity fraction (cases divided by the total number of tests) instead.
As in the other figures,  we use three credibility levels in Figure~\ref{fig:combined_forecast_plot}, and observed values are displayed as black circles.
Our one-week ahead probabilistic forecasts for both observed deaths in the upper half of Figure~\ref{fig:combined_forecast_plot} generally capture the observed values, indicating our method to be precise and well calibrated.
The four-week ahead forecasts predict the data well in times of relative stability, but exhibit poor performance when time-varying parameters are changing rapidly, such as during the winter wave.
In this case, when forecasting four weeks out, we tend to underestimate the rise at the beginning of the wave and overestimate the fall at the end of the wave.
This is not of major concern because the four-week time horizon is long enough that interventions and behavioral changes may take effect that are not foreseen by the model.
Scenario-based modeling, where some values are specified for the future time-varying parameters, rather than simulating from the prior, may be more appropriate for this task.
\par
\begin{figure}
    \centering
    \includegraphics[width=0.75\textwidth]{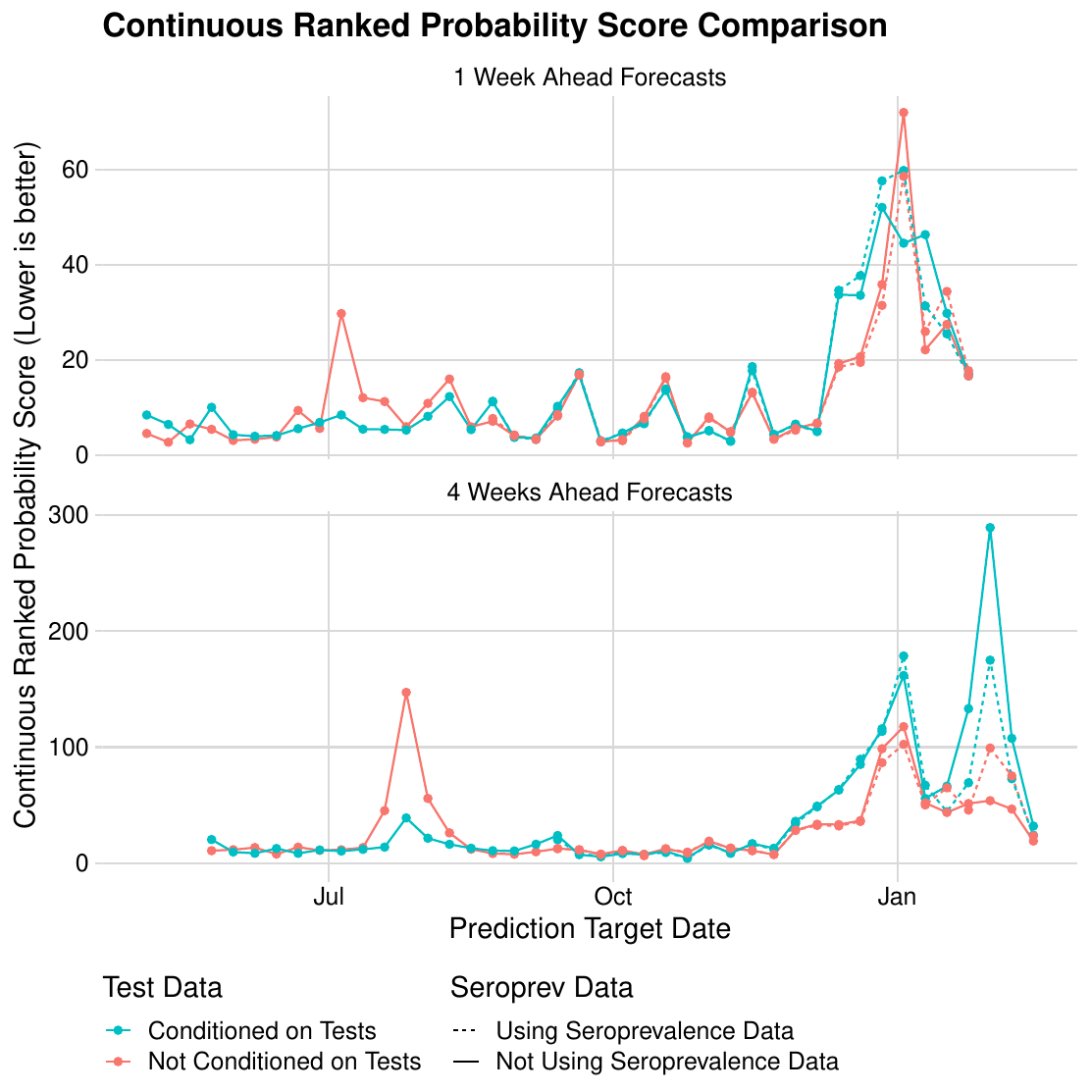}
    \caption{Comparison of Continuous Rank Probability Score for models fit to the Orange County data.
    Lower is better.}
    \label{fig:forecast_crps_plot}
\end{figure}

We now compare our forecasting results to three variants of our model, which make use of different data streams.
Each model can either be conditioned on tests or not conditioned on tests.
When conditioning on tests, we use the case emission model given by \eqref{eqn:caseemission}.
When not conditioning on tests, we use the case emission model given by \eqref{eqn:altcaseemission}.
Each model can also make use of the seroprevalence data or not.
When using the seroprevalence data, we use the emission distribution given by \eqref{eqn:seroprevemission}.
When not using the seroprevalence data, no emission distribution is used.
The model results discussed above are for the model which is conditioned on tests and uses the seroprevalence data.
We compare these models by calculating the Continuous Ranked Probability Score (CRPS) \citep{CRPS}, as implemented in the \texttt{scoringRules} package \citep{scoringRules}, facilitated by the \texttt{scoringutils} package \citep{scoringutils}.
We present comparisons of these scores in Figure~\ref{fig:forecast_crps_plot}.
We only show scores based on deaths because comparisons based on cases would require developing a method to forecast future test counts, which we have not considered in this work.

From Figure~\ref{fig:forecast_crps_plot}, we observe that all models appear to perform similarly throughout much of the assessed time period.
However, in both one week and four week ahead forecasts, the models which are conditioned on tests tend to score slightly better in the late summer period, when testing policy was rapidly changing.
During the winter surge, the differences in the model forecasting abilities are more pronounced, with the models not conditioned on tests appearing to be consistently superior to those which are conditioned on tests.
There is no clear pattern differentiating the models which use the seroprevalence data from those that do not.

\section{Discussion}
\label{sec:discussion}
We developed a Bayesian SARS-CoV-2 transmission model that integrates information from incidence, mortality, and seroprevalence data.
Our approach combines an ODE-based SEIR compartmental model of SARS-CoV-2 transmission dynamics and a carefully constructed surveillance model for cases,  deaths, and seroprevalence.
Importantly, our method accounts for variability in the number of SARS-CoV-2 diagnostics tests across time, thus ensuring that we do not confuse increases in testing with increases in incidence.
Another distinguishing feature of our approach is nonparametric modeling of changes in key transmission and surveillance model parameters.
Since we are integrating multiple sources of information, we can afford to be fairly ambitious and to include three such parameters into our model.
Changes in one of these parameters accounts for changing the strength of preferentially testing SARS-CoV-2 infected individuals, which helps us avoid an important source of potential bias when inferring transmission model parameters.
We reconstruct latent dynamics of the two pre-delta COVID-19 waves in Orange County, CA and estimate that 32--72\% of Orange County residents experienced SARS-CoV-2 infection by mid-January 2021.
Retrospective analysis shows that our model produces accurate and well calibrated one week ahead and reasonable four week ahead forecasts, but the latter lack accuracy during periods when SARS-CoV-2 transmission dynamics and mitigation policies change rapidly.
Additionally, we evaluated our forecasting performance when including or excluding certain data streams (test counts and seroprevalence study data) and found that incorporating negative test counts into our model was useful near the beginning of the modeling period, when testing policy was changing rapidly, but less useful when the policy became consistent.
We also found that excluding seroprevalence data did not negatively affect our forecasting ability.

Our primary focus in this work was on developing a framework for integrating multiple data streams into a transmission model.
However, there are a number of extensions we could pursue to improve the realism of the assumed transmission dynamics and strengthen the model's forecasting skill.
Our model assumes that the population of interest is well mixed and that all individuals in the population infect others and get infected at the same per capita rate.
In fact, the actual SARS-CoV-2 transmission process is much more complex because individuals come into contact with each other based on their geographical and social network proximity.
Furthermore, it is well established that COVID-19 disease progression process depends on the individual's age and other characteristics \citep{kim2020risk,bhargava2020predictors, petrilli2020factors}.
Similarly, all transmission model parameters may depend on the vaccination status of an individual.
Fortunately, compartmental models can be extended to account for these complexities.
For example, we can stratify each model compartment by age, vaccination status, and geographical location, as is commonly done in epidemiological modeling \citep{li2008continuous, van2008spatial}.
\par
We have addressed changes in control/mitigation measures and in human behavior by nonparametrically modeling variability of some of the SARS-CoV-2 transmission model parameters across time.
\citet{anderson2020} and \citet{miller2020} use parametric approaches to model the effects of mitigation measures on $R_0$.
It would be interesting to try a semi-parametric approach that combines parametric and non-parametric components, which would allow us to include indicators of human behavior (e.g., mobility data as in \citet{miller2020}) into our inference and forecasting.
\par
In this paper, we have sidestepped the thorny issue of reporting delays by restricting our analyses to time periods in which the data have stabilized.
Hence, our analyses should be robust to reporting delays so long as we have either ``run out the clock" on the extent of the delays or reporting delays do not differ between positive and negative COVID-19 diagnostic tests. A useful set of extensions that would make our model more useful for real-time surveillance involve estimating the reporting delay distribution \citep{hohle2014bayesian,stoner2019multivariate} and using this distribution in our surveillance model.
\par
Finally, we would like to point out that our deterministic representation of the latent epidemic process could be substituted for a fully stochastic model where the latent epidemic is represented as a Markov jump process, albeit with some loss of computational efficiency. In our large population setting, this could be achieved via simulation-based methods \citep{breto2009time, andrieu2010particle,dukic2012tracking}, data augmentation \citep{pooley2015using,nguyen2021stochastic}, or a variety of approximations of the latent stochastic epidemic process \citep{lekone2006statistical,cauchemez2008likelihood,fintzi2020linear}.
Scaling our model to the state or national level could be done by analyzing multiple counties independently or by building a Bayesian hierarchical model that would allow borrowing information among counties.
An even more ambitious undertaking would be allowing importation/exportation events across county lines, as was done by \citet{pei2021burden}. We hope that our methodology and other works in this spirit, along with better quality of surveillance data, will provide us with better predictive analytics tools when the next pandemic strikes.

\section*{Acknowledgements}
This work utilized the infrastructure for high-performance and high-throughput computing, research data storage and analysis, and scientific software tool integration built, operated, and updated by the Research Cyberinfrastructure Center (RCIC) at the University of California, Irvine.
We are grateful for funding from the UCI Infectious Disease Science Initiative.
This work was made possible in part through support from the UC CDPH Modeling Consortium.
D.B, I.H.G, and V.M.M were supported in part by NIH grant R01AI147336.
V.M.M was supported in part by NIH grant R01AI170204 and NSF grant DMS 1936833.
ER was supported by the Division of Intramural Research, NIAID, NIH.

\section*{Disclaimers}
This project has been funded in part with federal funds from the National Cancer Institute, National Institutes of Health, under Contract No. 75N91019D00024, Task Order No. 75N91019F00130.
This work was in part supported by the intramural research programs of the National Institutes of Health, Bethesda, MD.
The content of this publication does not necessarily reflect the views or policies of the Department of Health and Human Services, nor does mention of trade names, commercial products, or organizations imply endorsement by the U.S. Government.
\clearpage
\bibliographystyle{Apalike-JASA}
\bibliography{oc_report}

\clearpage
\appendix
\singlespacing

\setcounter{page}{1}
\setcounter{table}{0}

\setcounter{equation}{0}
\setcounter{section}{0}
\setcounter{figure}{0}

\def\thesection{S-\arabic{section}}
\renewcommand\thefigure{\thesection\-.\arabic{figure}}
\renewcommand\thetable{\thesection\-.\arabic{table}}

\begin{center}
\Large Supplementary Materials for ``Semi-parametric modeling of SARS-CoV-2 transmission using tests, cases, deaths, and seroprevalence data''
\end{center}

\section{Simulation study}

\begin{figure}[h]
    \centering
    \includegraphics[width=1.0\textwidth]{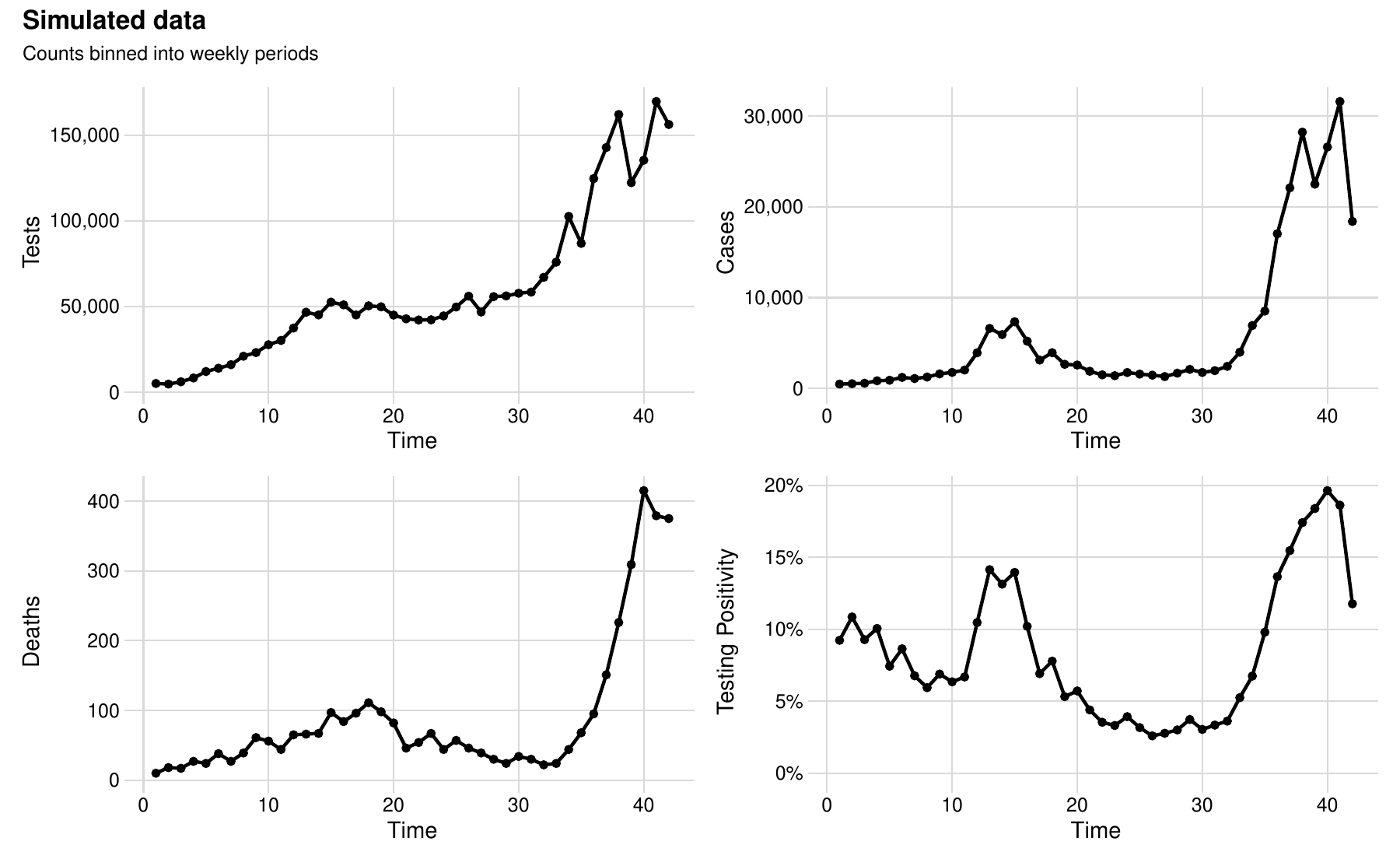}
    \caption{
    Simulated data.
    The figure shows weekly counts of tests, cases (positive tests), reported deaths due to COVID-19, as well testing positivity.}
    \label{fig:simulated_binned_data_plot}
\end{figure}

\begin{figure}
    \centering
    \includegraphics[width=1.0\textwidth]{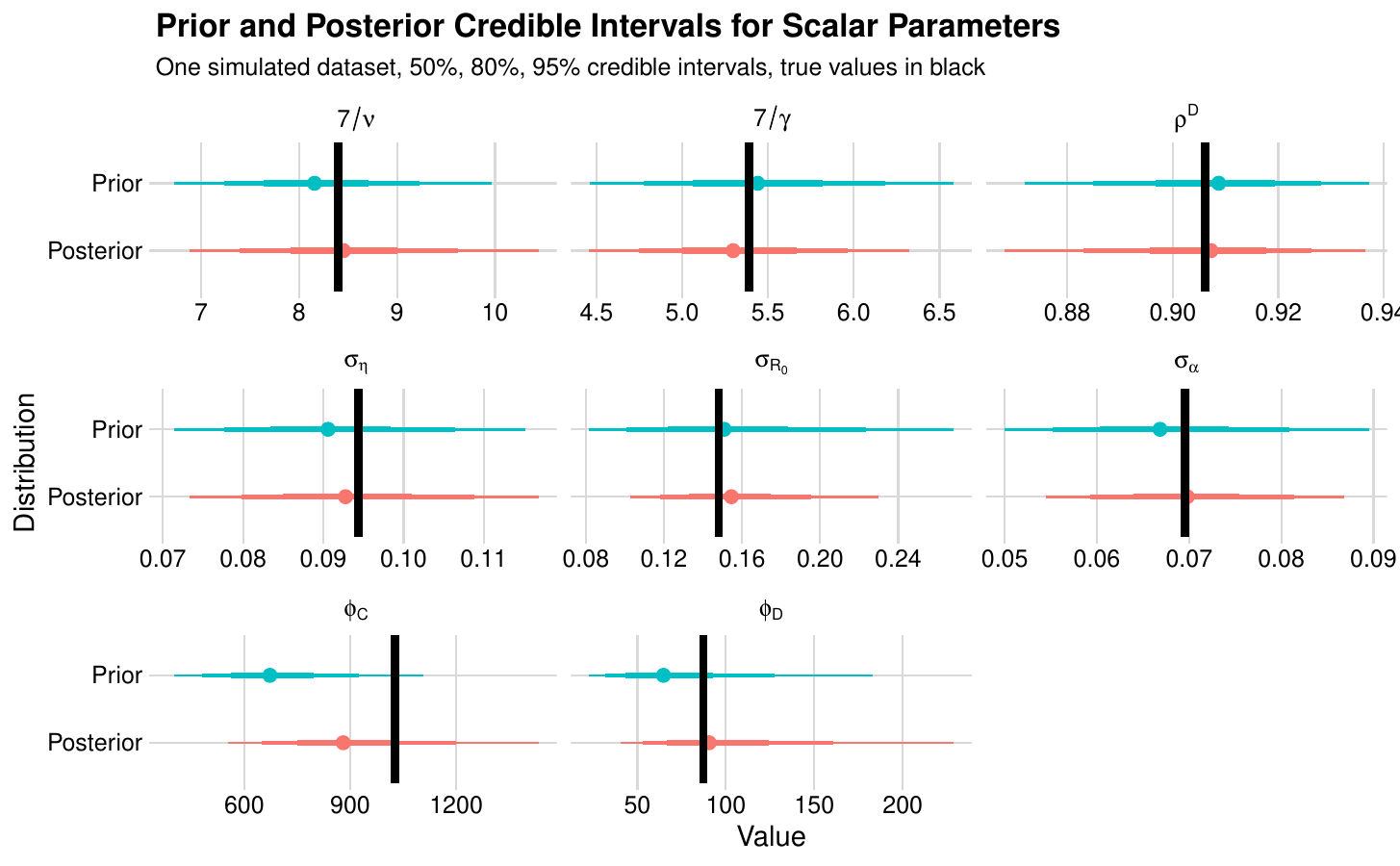}
    \caption{Prior and posterior credible intervals for scalar parameters for a model fit to the dataset presented in Figure~\ref{fig:simulated_binned_data_plot}.
    True values for the simulated parameters are indicated by solid black lines.}
    \label{fig:single_generated_quantities_simulation_scalar_plot}
\end{figure}

\begin{figure}
    \centering
    \includegraphics[width=1.0\textwidth]{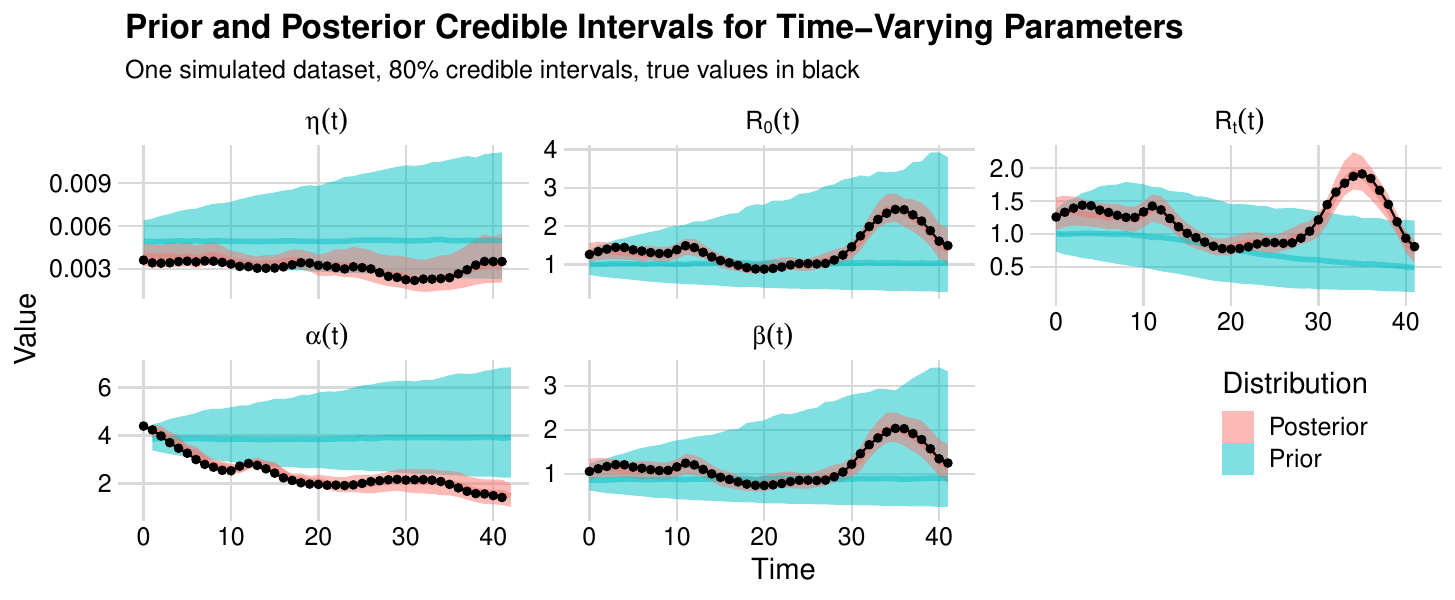}
    \caption{Prior and posterior 80\% credible intervals for time-varying parameters for a model fit to the dataset presented in Figure~\ref{fig:simulated_binned_data_plot}.
    True values for the simulated parameters are indicated by solid black lines.}
    \label{fig:single_generated_quantities_simulation_time_varying_plot}
\end{figure}

\begin{figure}
    \centering
    \includegraphics[width=1.0\textwidth]{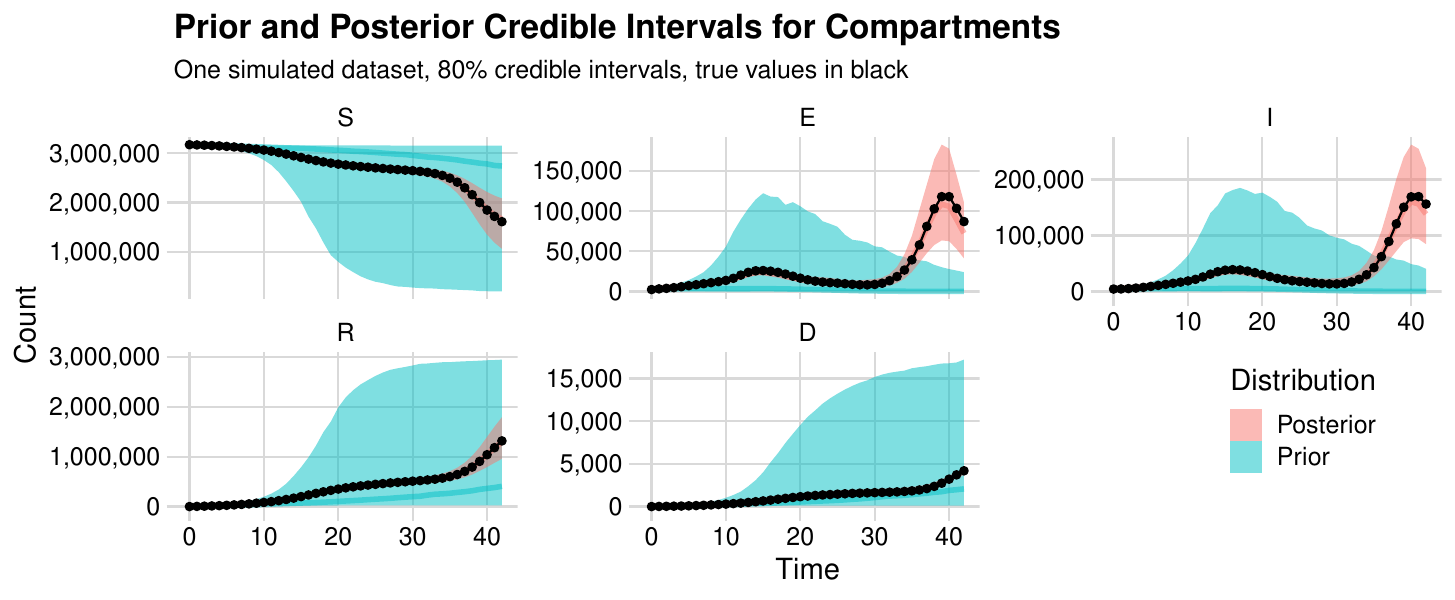}
    \caption{Prior and posterior 80\% credible intervals for latent compartments for a model fit to the dataset presented in Figure~\ref{fig:simulated_binned_data_plot}.
    True values for the simulated compartment sizes are indicated by solid black lines.}
    \label{fig:single_generated_quantities_simulation_compartment_plot}
\end{figure}

\begin{figure}
    \centering
    \includegraphics[width=1.0\textwidth]{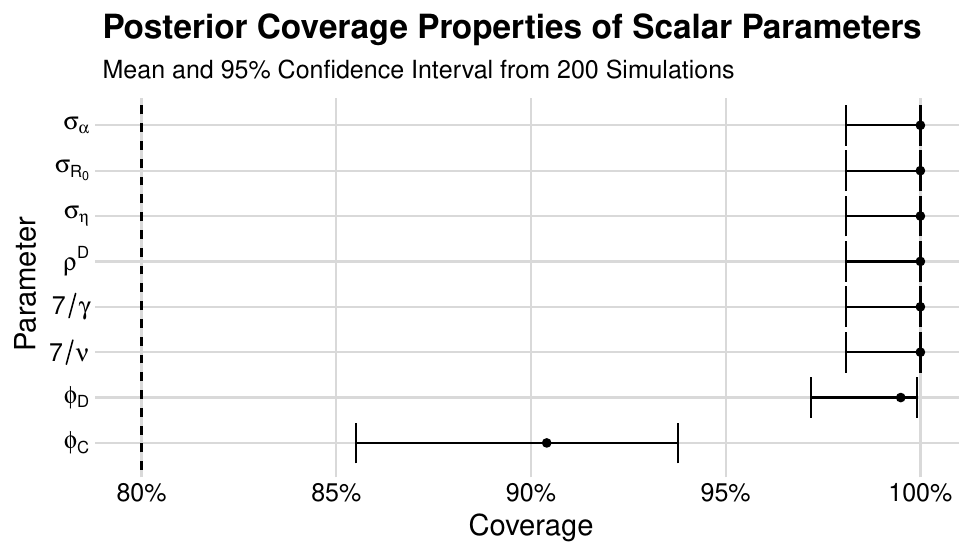}
    \caption{Coverage properties of 80\% posterior credible intervals for scalar parameters from models fit to 200 simulated datasets.
    Nominal coverage is indicated by the dashed line.}
    \label{fig:generated_quantities_simulation_scalar_coverage_plot}
\end{figure}

\begin{figure}
    \centering
    \includegraphics[width=1.0\textwidth]{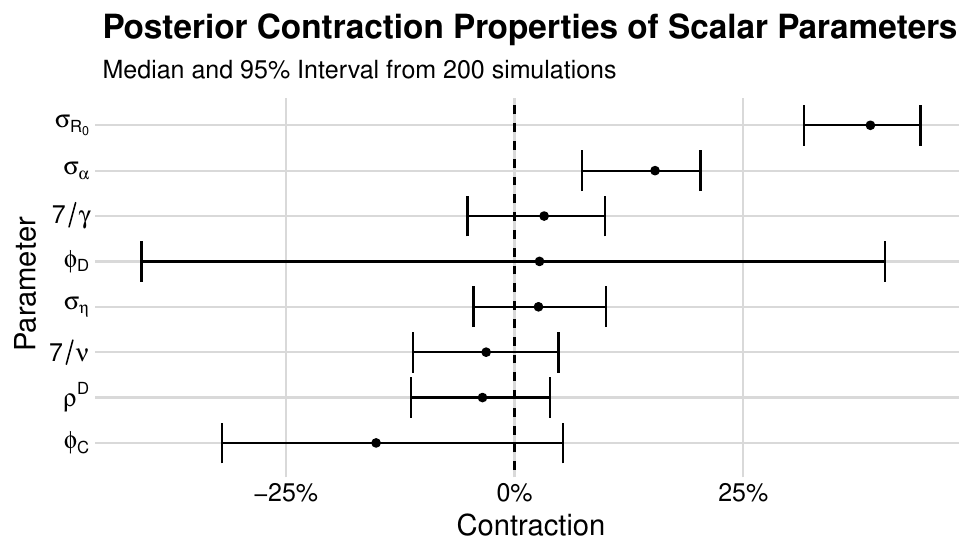}
    \caption{Contraction properties of scalar parameters from models fit to 200 simulated datasets.
    Contraction is calculated as one minus the ratio of standard deviation of the posterior and the prior.}
    \label{fig:generated_quantities_simulation_scalar_shrinkage_plot}
\end{figure}

\begin{figure}
    \centering
    \includegraphics[width=1.0\textwidth]{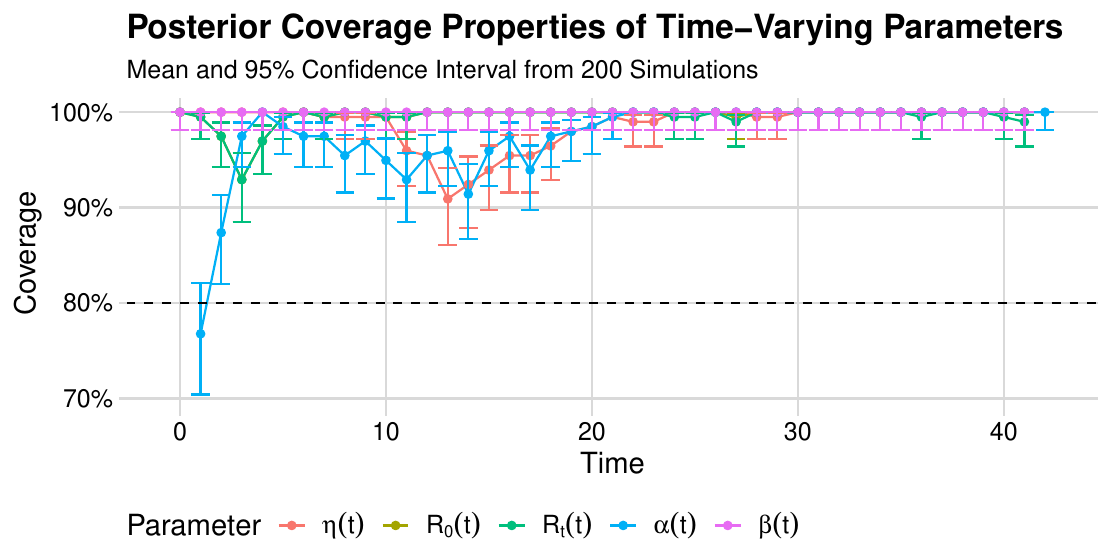}
    \caption{Coverage properties of 80\% posterior credible intervals for time-varying parameters from models fit to 200 simulated datasets.
    Nominal coverage is indicated by the dashed line.}
    \label{fig:generated_quantities_simulation_time_varying_coverage_plot}
\end{figure}

\begin{figure}
    \centering
    \includegraphics[width=1.0\textwidth]{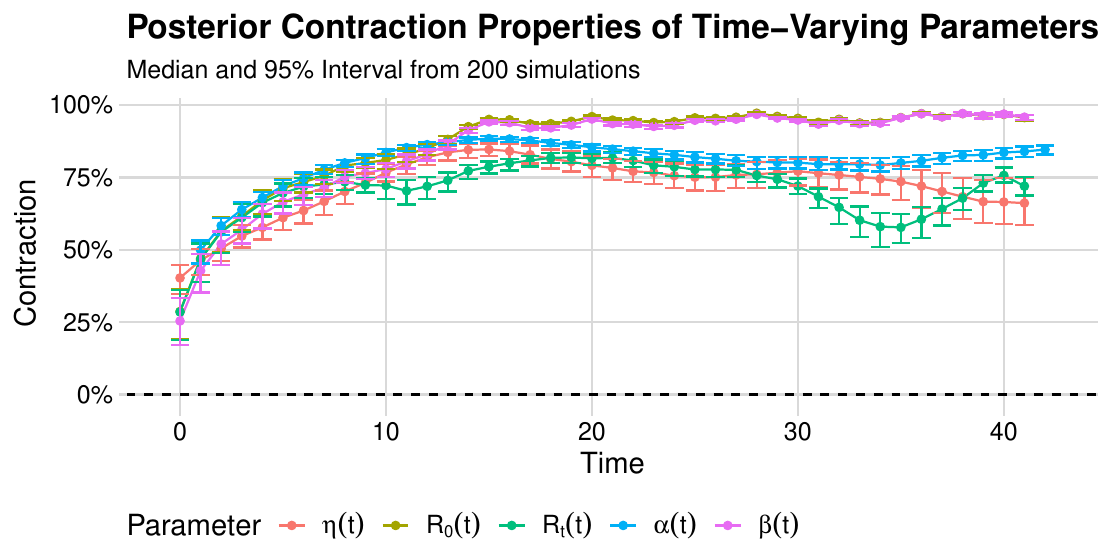}
    \caption{Contraction properties of time-varying parameters from models fit to 200 simulated datasets.
    Contraction is calculated as one minus the ratio of standard deviation of the posterior and the prior.}
    \label{fig:generated_quantities_simulation_time_varying_shrinkage_plot}
\end{figure}

We performed a simulation study on 200 datasets to validate our models.
We use the same prior distributions for the parameters as in the main text.
These distributions are presented in Table~\ref{table:simulation_parameters}.
We purposely chose parameter values that resulted in data similar to the Orange Country data used in the main text.
Exact values for these parameters are presented in Table~\ref{table:simulation_parameters}.
One of the 200 simulated datasets is presented in Figure~\ref{fig:simulated_binned_data_plot}.
Figures~\ref{fig:single_generated_quantities_simulation_scalar_plot}--\ref{fig:single_generated_quantities_simulation_compartment_plot} present the prior and posterior distribution for this single dataset.
Figures~\ref{fig:generated_quantities_simulation_scalar_coverage_plot}--\ref{fig:generated_quantities_simulation_time_varying_shrinkage_plot} show coverage and contraction properties for the whole simulation study.
Contraction is calculated as one minus the ratio of standard deviation of the posterior and the prior.
Commentary on these results is presented in Section~\ref{sec:results} of the main text.

\begin{table}
	\caption[Parameters and priors.]{Model parameters and their prior distributions.}
	\label{table:priors}
	\scriptsize\centering
	\begin{tabularx}{\textwidth}{cXllc}
	\thead{Parameter} & \thead{Interpretation} & \thead{Prior} & \thead{Prior Median\\ (95\% Interval)} & \thead{Source} \\ \hline
$S_0$ & Initial susceptible proportion & Logit-Normal(6, 0.25) & \makecell{0.998 \\ (0.993, 0.999)} & \\
$\tilde{I}_{0}$ & Initial proportion of non-susceptibles who are infectious & Logit-Normal(0.6, 0.0009) & \makecell{0.646 \\ (0.632, 0.659)} & \\
$\exp\left(\tilde{R}_{0,1}\right)$ & Initial basic reproduction number & Log-Normal(0, 0.0625) & \makecell{1.000 \\ (0.613, 1.630)} & \\
$1 / \gamma$ & Mean latent period (weeks) & Log-Normal(-0.25, 0.01) & \makecell{0.779 \\ (0.640, 0.947)} & \cite{Xin2021} \\
$1 / \nu$ & Mean infectious period (weeks) & Log-Normal(0.15, 0.01) & \makecell{1.160 \\ (0.955, 1.410)} & \cite{Byrnee039856} \\
$\expit\left(\tilde{\eta}_1\right)$ & Initial infection fatality ratio & Logit-Normal(-5.3, 0.04) & \makecell{0.00497 \\ (0.00336, 0.00733)} & \cite{Bruckner2021} \\
$\rho^D$ & Mean death detection rate & Logit-Normal(2.3, 0.04) & \makecell{0.909 \\ (0.871, 0.937)} & \cite{Bruckner2021} \\
$\phi_D$ & over-dispersion in observed deaths Negative-Binomial model & Log-Normal(4.16, 0.293) & \makecell{ 63.9 \\ ( 22.1, 185.0)} & \\
$\exp\left(\tilde{\alpha}_1\right)$ & Initial proportion in proportional odds test positivity model & Log-Normal(1.35, 0.0121) & \makecell{3.86 \\ (3.11, 4.79)} & \\
$\phi_C$ & over-dispersion in observed cases beta-binomial model & Log-Normal(6.5, 0.0673) & \makecell{ 665 \\ (400, 1110)} & \\
$\sigma_{R_0}$ & Standard deviation of log-Guassian Markov random field for time-varying $R_0$ & Log-Normal(-1.9, 0.09) & \makecell{0.1500 \\ (0.0831, 0.2690)} & \\
$\sigma_\eta$ & Standard deviation of logit-Guassian Markov random field for time-varying $\eta$ & Log-Normal(-2.4, 0.0144) & \makecell{0.0907 \\ (0.0717, 0.1150)} & \\
$\sigma_\alpha$ & Standard deviation of log-Guassian Markov random field for time-varying $\alpha$ & Log-Normal(-2.7, 0.0225) & \makecell{0.0672 \\ (0.0501, 0.0902)} & \\
\hline
$\tilde{\rho}^Y_1$ & Initial case detection rate & Logit-Normal(-2.5, 0.01) & \makecell{0.0759 \\ (0.0632, 0.0908)} & \cite{Bruckner2021} \\
$\phi_Y$ & over-dispersion in observed cases Negative-Binomial model & Log-Normal(3.93, 0.0684) & \makecell{51.1 \\ (30.6, 85.3)} & \\
$\sigma_{\rho^Y}$ & Standard deviation of logit-Guassian Markov random field for time-varying $\rho^Y$ & Log-Normal(-2.2, 0.04) & \makecell{0.1110 \\ (0.0749, 0.1640)} &
	\end{tabularx}
\end{table}

\begin{table}%[htbp]
	\caption[Simulation parameters.]{Simulation parameters.}
	\label{table:simulation_parameters}
	\scriptsize\centering
	\begin{tabularx}{\textwidth}{cXllc}
	\thead{Parameter} & \thead{Interpretation} & \thead{Value}  \\ \hline
$S_0$ & Initial susceptible proportion & 0.9979 \\
$\tilde{I}_{0}$ & Initial proportion of non-susceptibles who are infectious & 0.6455 \\
$\exp\left(\tilde{R}_{0,1}\right)$ & Initial basic reproduction number & 1.2602 \\
$1 / \gamma$ & Mean latent period (weeks) &  0.7697 \\
$1 / \nu$ & Mean infectious period (weeks) & 1.1997 \\
$\expit\left(\tilde{\eta}_1\right)$ & Initial infection fatality ratio & 0.0005 \\
$\rho^D$ & Mean death detection rate &  0.9061 \\
$\phi_D$ & over-dispersion in observed deaths Negative-Binomial model & 87.2776 \\
$\exp\left(\tilde{\alpha}_1\right)$ & Initial proportion in proportional odds test positivity model & 4.3958 \\
$\phi_C$ & over-dispersion in observed cases Beta-Binomial model & 1,026.6765 \\
$\sigma_{R_0}$ & Standard deviation of log-Guassian Markov random field for time-varying $R_0$ & 0.1481 \\
$\sigma_\eta$ & Standard deviation of logit-Guassian Markov random field for time-varying $\eta$ & 0.0944 \\
$\sigma_\alpha$ & Standard deviation of log-Guassian Markov random field for time-varying $\alpha$ & 0.0696 \\
\hline
	\end{tabularx}
\end{table}

\begin{figure}
    \centering
    \includegraphics[width=1.0\textwidth]{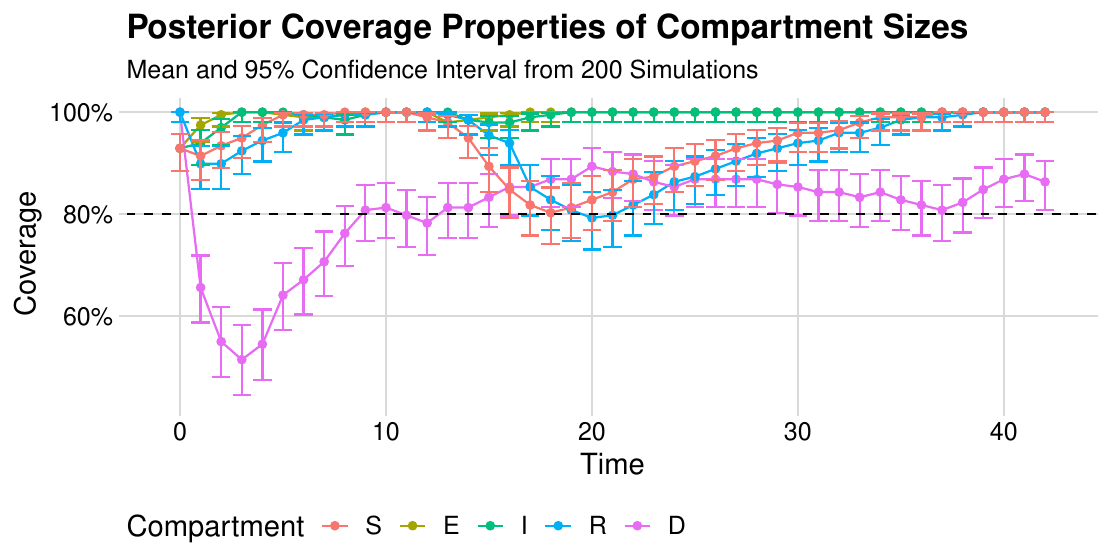}
    \caption{Coverage properties of 80\% posterior credible intervals for latent compartments from models fit to 200 simulated datasets.
    Nominal coverage is indicated by the dashed line.}
    \label{fig:generated_quantities_simulation_compartment_coverage_plot}
\end{figure}

\begin{figure}
    \centering
    \includegraphics[width=1.0\textwidth]{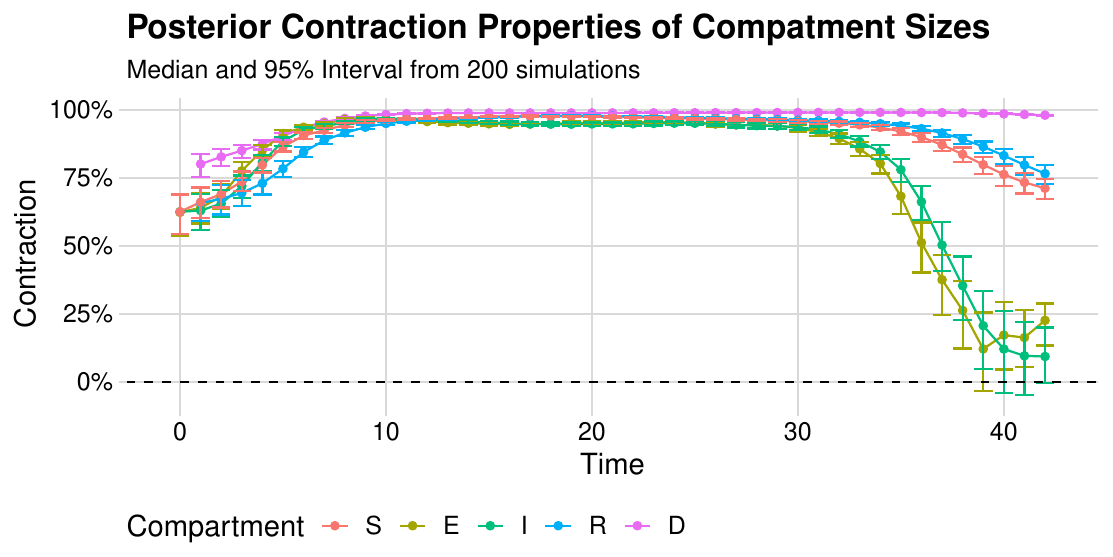}
    \caption{Contraction properties of latent compartments from models fit to 200 simulated datasets.
    Contraction is calculated as one minus the ratio of standard deviation of the posterior and the prior.}
    \label{fig:generated_quantities_simulation_compartment_shrinkage_plot}
\end{figure}

\section{Comparison with \texttt{epidemia}}
\label{epidemia_section}

We used the \texttt{epidemia} R package to infer \( R_t \) in the same 200 simulated datasets, as well as the Orange County data. 
Statistical details of these methods are presented below.
Commentary on these results is presented in Section~\ref{sec:results} of the main text.

The \texttt{epidemia} package can be used to create different branching process inspired models to estimate the effective reproduction number. In contrast to the compartmental model used in this paper, branching process inspired models have related the mean of current incidence to a weighted sum of previous incidence and the effective reproduction number $R_{t}$. Let $I_{t}$ be the incidence at time $t$, $R_{t}$ be the effective reproduction number at time $t$, and $g(t)$ be the probability density function of the generation time distribution (the time between an individual becoming infected and infecting another individual; under the compartmental model framework this is usually taken to be equivalent to the sum of the latent period and the infectious period). Then the mean relationship used is:
\begin{equation*}
    E[I_{t}|I_{1}, \dots, I_{t-1}] = R_{t}\sum_{s=1}^{t-1}I_{s}g(t-s).
\end{equation*}
For the model we used in this study, we then added an observation model for new cases, modeled the effective reproduction number as a random walk, and modeled unobserved incidence as an autoregressive normal random variable with variance equal to the mean multiplied by an over-dispersion parameter.
 \begin{align*}
\tau &\sim \text{exp}(\lambda)  \quad \quad \text{Hyperprior for unobserved incidence}\\
I_{\nu} &\sim \text{exp}(\tau) \quad \quad \text{Prior on unobserved incidence $\nu$ days before observation}\\
I_{\nu+1}, \dots, I_{0} &= I_{\nu}  \quad \quad \text{Unobserved incidence}\\
    \sigma & \sim \text{Truncated-Normal}(0, 0.1^{2}) \\
   \log{R_{0}} &\sim \text{Normal}(\log{2}, 0.2^{2})\quad \quad \text{Prior on $R_{0}$} \\
    \log{R_{t}}|\log{R_{t-1}} &\sim \text{Normal}(\log{R_{t-1}}, \sigma)  \quad \quad \text{Random Walk prior on $R_{t}$}\\
\psi &\sim \text{Normal}(10,2) \quad \quad \text{Prior on variance parameter for incidence} \\
I_{t}|I_{\nu}, \dots, I_{t-1} &\sim \text{Normal}(R_{t}\sum_{s<t}I_{s}g_{t-s}, \psi) \quad \quad \text{Model for incidence} \\
    \alpha &\sim \text{Normal}(0.13, 0.7^2)
\quad \quad \text{Prior on case detection rate} \\
y_{t} &= \alpha_{t}\sum_{s<t}I_{s}\pi_{t-s} \quad \quad \text{Mean of observed data model}\\
\phi & \sim P(\phi) \quad \quad \text{Prior on dispersion parameter for observed data} \\
Y_{t} &\sim \text{Neg-Binom}(y_{t}, \phi) \quad \quad \text{Observed data model}\\
\end{align*}
Here $\pi_{t}$ are the values of the probability density function for the delay distribution, the time between an individual being infected and being observed.
This distribution is assumed to be a gamma distribution with shape parameter one and mean equal to the true mean latent period. 
To sample from the posterior distribution, \texttt{epidemia} uses Hamiltonian Monte Carlo via the \texttt{Stan} simulation software \citep{rstan}.
We draw 2000 posterior samples and discard the first 1000 for this analysis.

\begin{figure}
    \centering
    \includegraphics[width=1.0\textwidth]{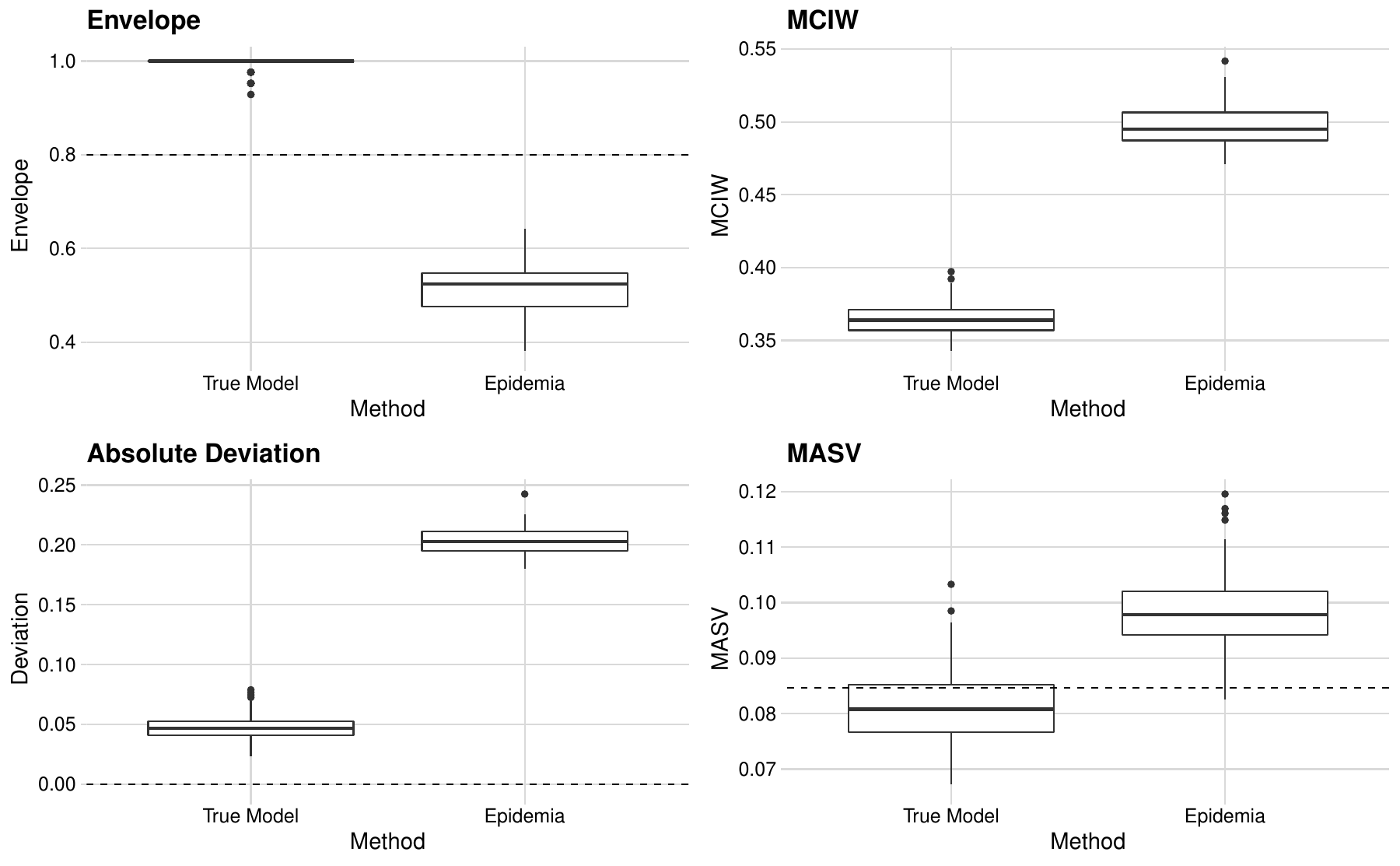}
    \caption{Properties of \( R_t \) estimation from 200 simulated data sets.
    The envelope is the proportion of time points which the 80\% posterior credible interval contains the true \( R_t \) value specified in the simulation.
    Mean credible interval width (MCIW) is the mean of credible interval widths across time points within a simulation replication.
    Absolute deviation is calculated as the mean of the absolute difference between the posterior median and the true \( R_t \) value at each time point.
    The mean absolute sequential variation (MASV) is the mean of the absolute difference between the posterior median at a time point and the posterior median at the previous time point.}
    \label{fig:rt_comparison_metrics_plot}
\end{figure}

\section{Comparison with modeling structured populations}
\label{sec:hetero}
Here, we demonstrate that semi-parametric modeling of key parameters can obviate the need for modeling heterogeneous populations with separate compartments.
We construct a model wherein a disease spreads among two subpopulations: the ``general" population and the ``vulnerable" population, which interact with each other.
Progression through compartments is governed by the following system of differential equations, with ``g" subscripts denoting the general subpopulation and ``v" subscripts denoting the vulnerable subpopulation and parameters having the same interpretations as in the main text.
The differential equations used for this model presented in \eqref{eqn:model_hetero_SIR_ODEs}.

\begin{equation}
\label{eqn:model_hetero_SIR_ODEs}
\begin{aligned}
\deriv{S_v}{t} &= -\left(\beta_{vv} I_v + \beta_{vg} I_g \right) \frac{S_v}{N} \\
\deriv{E_v}{t} &= \left(\beta_{vv} I_v + \beta_{vg} I_g \right) \frac{S_v}{N} - \gamma E_v \\
\deriv{I_v}{t} &= \gamma E_v - \nu I_v \\
\deriv{R_v}{t} &= \nu (1-\eta_v) I_v \\
\deriv{D_v}{t} &= \nu \eta_v I_v \\
\end{aligned}
\qquad \qquad \qquad
\begin{aligned}
\deriv{S_g}{t} &= -\left(\beta_{gg} I_g + \beta_{gv} I_v\right) \frac{S_g}{N} \\
\deriv{E_g}{t} &= \left(\beta_{gg} I_g + \beta_{gv} I_v \right) \frac{S_g}{N} - \gamma E_g \\
\deriv{I_g}{t} &= \gamma E_g - \nu I_g \\
\deriv{R_g}{t} &= \nu (1-\eta_g) I_g \\
\deriv{D_g}{t} &= \nu \eta_g I_g \\
\end{aligned}
\end{equation}

subject to initial conditions $ \bX(t_0) = \bx_0 $ and $ \bN(t_0) = \mathbf{0}$, where $\bx_0 = (S_{v0}, E_{v0}, I_{v0}, R_{v0}, D_{v0}, \linebreak[0] S_{g0}, E_{g0}, I_{g0}, R_{g0}, D_{g0})$ are initial compartment counts.

We only observed the unstratified case and death counts.
Observed cases and deaths are Poisson distributed with the rate parameter equal to the number of latent cases and deaths, respectively.

\[ Y_l \sim \text{Poisson}(\Delta N_{E_v I_v}(t_l) + \Delta N_{E_g I_g}(t_l)) \]
\[ M_l \sim \text{Poisson}(\Delta N_{I_v D_v}(t_l) + \Delta N_{I_g D_g}(t_l)) \]

\begin{figure}
    \centering
    \includegraphics[width=0.75\textwidth]{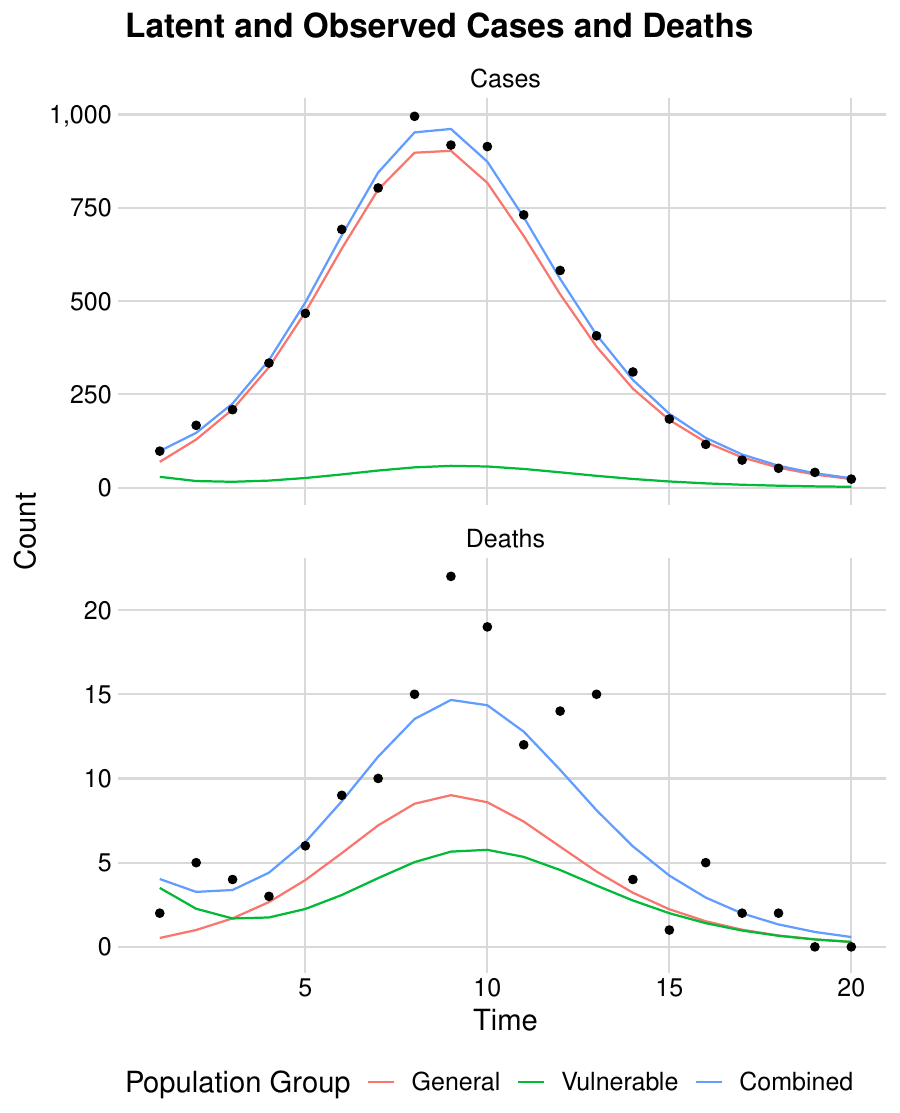}
    \caption{Latent new cases and deaths for vulnerable and general subpopulations, along with combined latent new cases and new deaths and observed (combined) new cases and new deaths for a simulated scenario.}
    \label{fig:data_ifr_age_structure_plot}
\end{figure}

We construct a scenario where a disease outbreak occurs in a small vulnerable population with a true infection-fatality ratio of 10\% before spreading to a larger general population with a true infection-fatality ratio of 1\%.
Because the outbreak spreads through the different populations at different times, the true population infection-fatality ratio varies in time.
Figure~\ref{fig:data_ifr_age_structure_plot} shows the latent new cases and new deaths for each subpopulation, as well as the combined latent new cases and new deaths, and the observed new cases and new deaths for this constructed scenario.

Now, we fit a semi-parametric model, similar to the one in the main text, to this data.
We model \( \eta(t) \) with a logit-Guassian Markov random field, as in the main text.
The differential equations used for this model presented in \eqref{eqn:model_nonparametric_SIR_ODEs}.
\begin{equation}
\label{eqn:model_nonparametric_SIR_ODEs}
\begin{aligned}
\deriv{S}{t} &= - \beta I \frac{S}{N}\\
\deriv{E}{t} &= \beta I \frac{S}{N} - \gamma E\\
\deriv{I}{t} &= \gamma E - \nu I\\
\deriv{R}{t} &= \nu (1 - \eta(t)) I\\
\deriv{D}{t} &= \nu \eta(t) I\\
\end{aligned}
\end{equation}

subject to initial conditions $ \bX(t_0) = \bx_0 $ and $ \bN(t_0) = \mathbf{0}$, where $\bx_0 = (S_0, E_0, I_0, R_0, D_0)$ are initial compartment counts.

\begin{figure}
    \centering
    \includegraphics[width=0.75\textwidth]{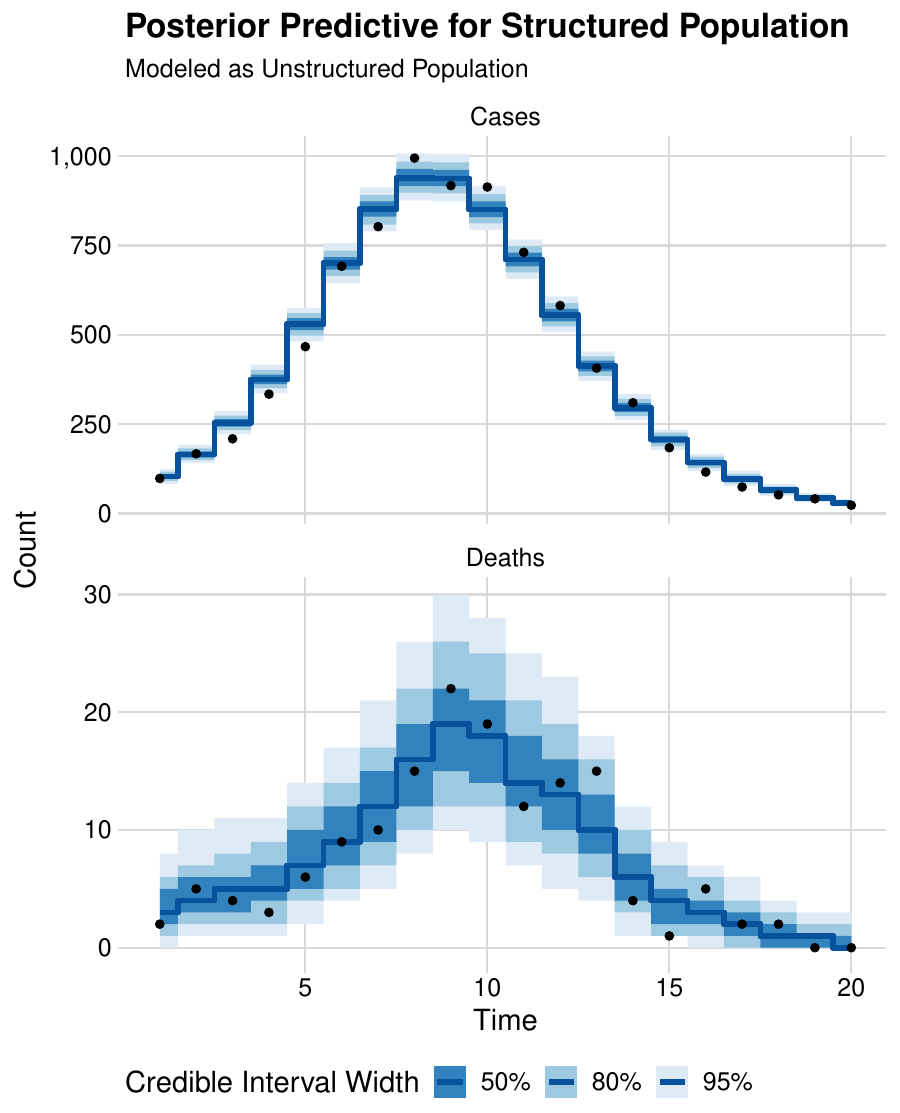}
    \caption{Posterior predictive distributions for a model with non-parametric IFR fit to a simulated dataset with a heterogeneous population.
    The case and death data used are shown as black dots.}
    \label{fig:posterior_predictive_ifr_age_structure_plot}
\end{figure}

\begin{figure}
    \centering
    \includegraphics[width=0.75\textwidth]{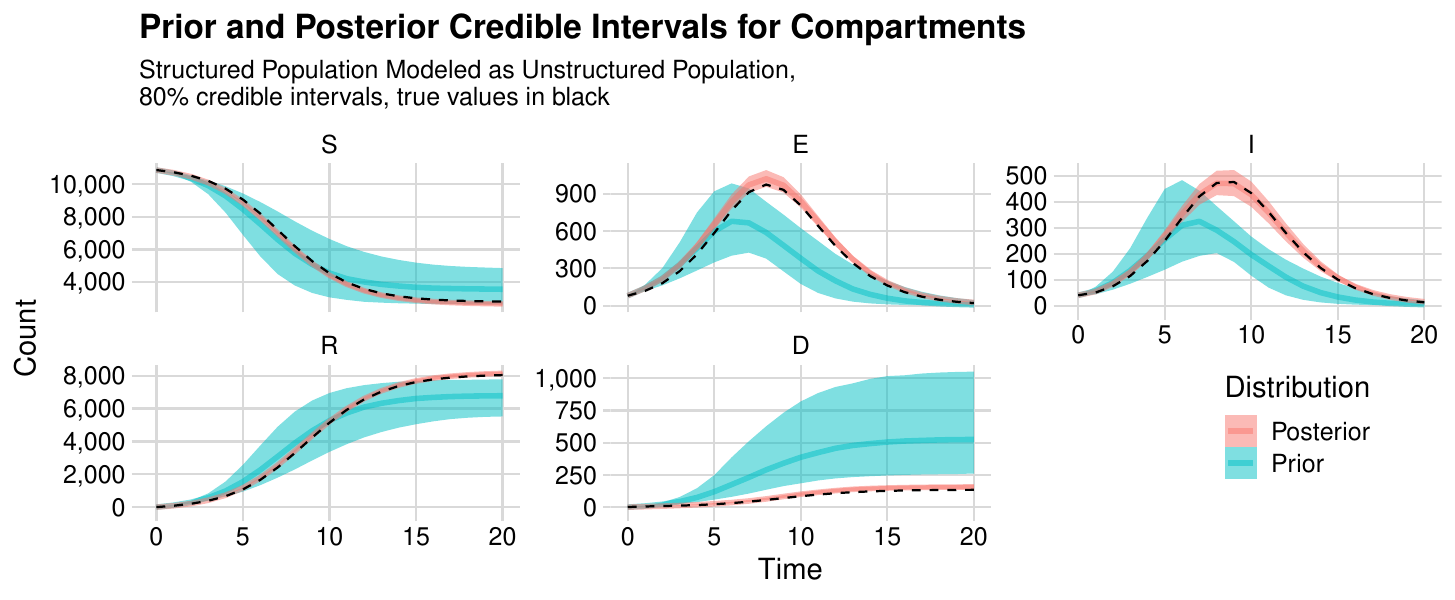}
    \caption{Prior and Posterior distributions for latent compartments for a model with non-parametric IFR fit to a simulated dataset with a heterogeneous population.
    The true time-varying parameters are indicated by the dashed line.}
    \label{fig:ifr_age_structure_generated_quantities_simulation_compartment_plot}
\end{figure}

\begin{figure}
    \centering
    \includegraphics[width=0.75\textwidth]{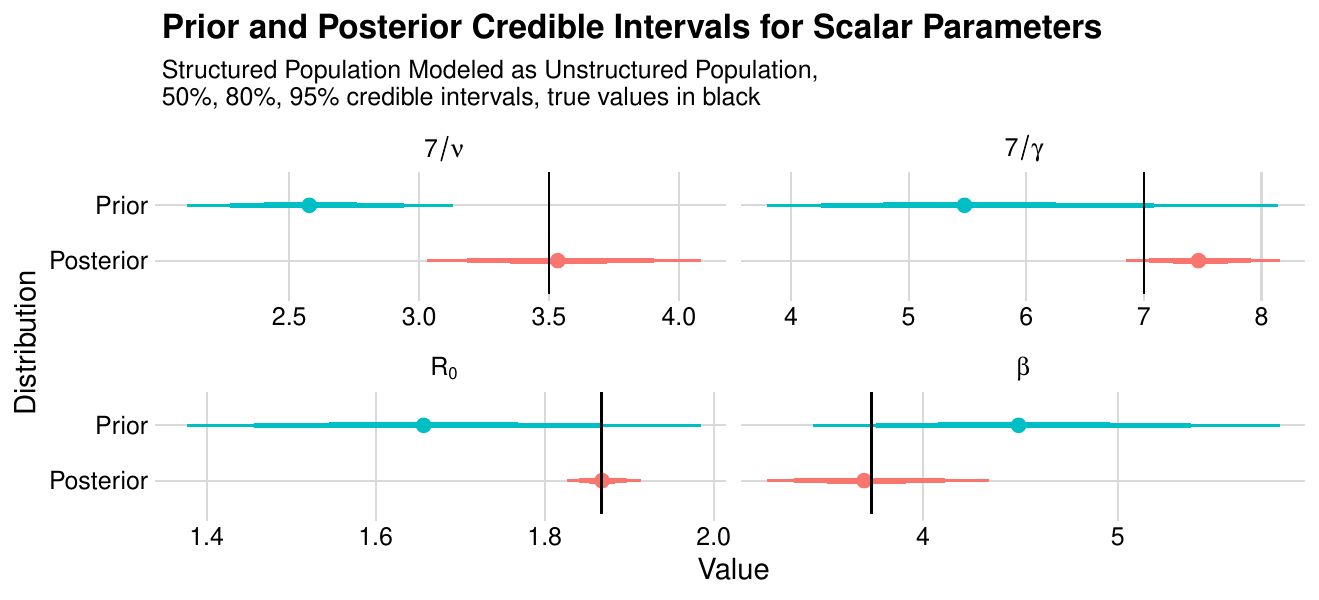}
    \caption{Prior and posterior distributions for scalar parameters for a model with non-parametric IFR fit to a simulated dataset with a heterogeneous population.
    The true time-varying parameters are indicated by the vertical line.}
    \label{fig:ifr_age_structure_generated_quantities_simulation_scalar_plot}
\end{figure}

\begin{figure}
    \centering
    \includegraphics[width=0.75\textwidth]{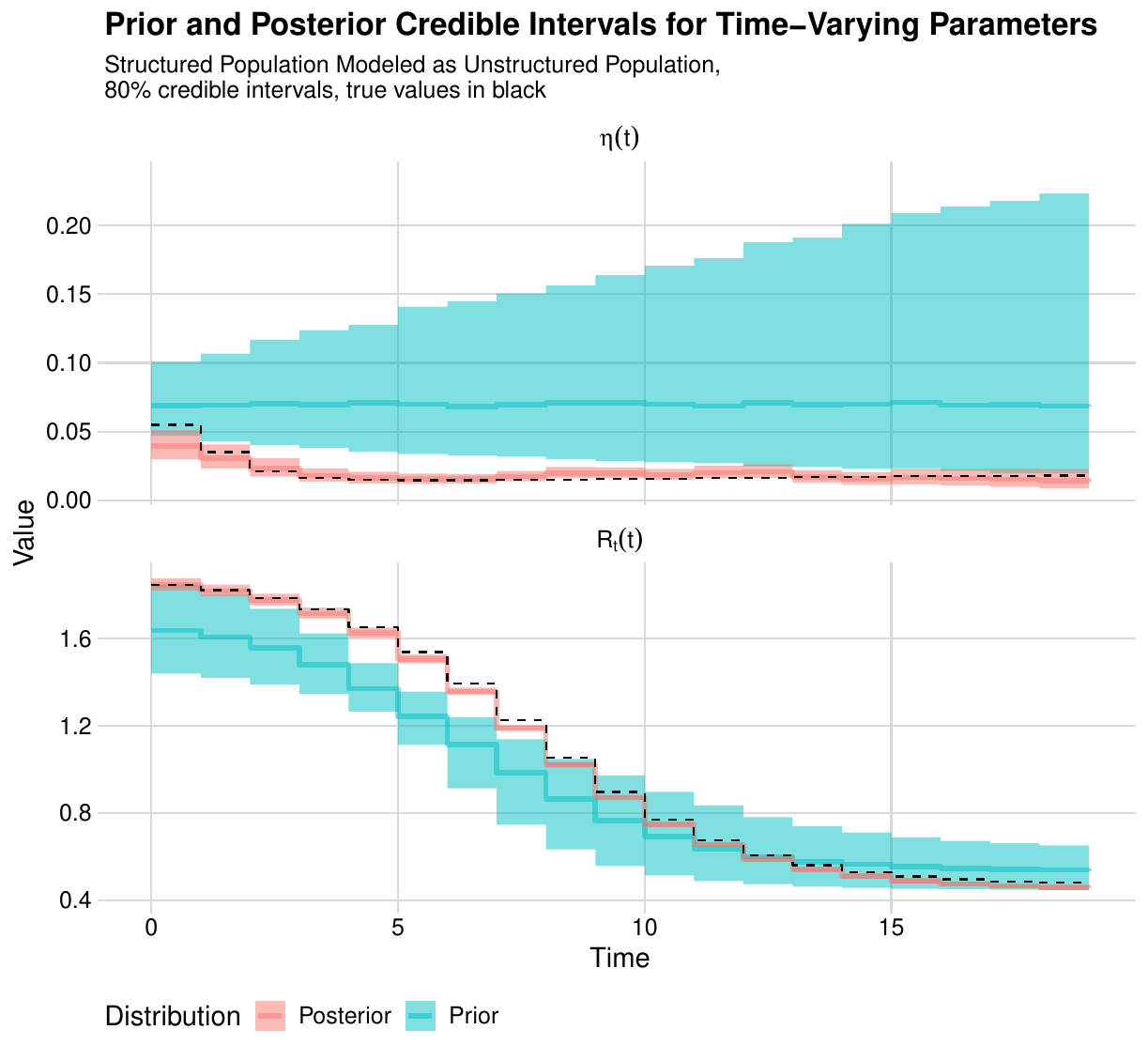}
    \caption{Prior and Posterior distributions for time-varying parameters for a model with non-parametric IFR fit to a simulated dataset with a heterogeneous population.
    The true time-varying parameters are indicated by the dashed line.}
    \label{fig:ifr_age_structure_generated_quantities_simulation_time_varying_plot}
\end{figure}

Figures~\ref{fig:posterior_predictive_ifr_age_structure_plot}--\ref{fig:ifr_age_structure_generated_quantities_simulation_time_varying_plot}, demonstrate, that when we fit our semi-parametric model, we can generally fit the data well and recover the true values of the parameters without modeling the two heterogeneous populations.

\section{Sensitivity analysis}
\label{sec:sensitivity}
\begin{figure}
    \centering
    \includegraphics[width=1.0\textwidth]{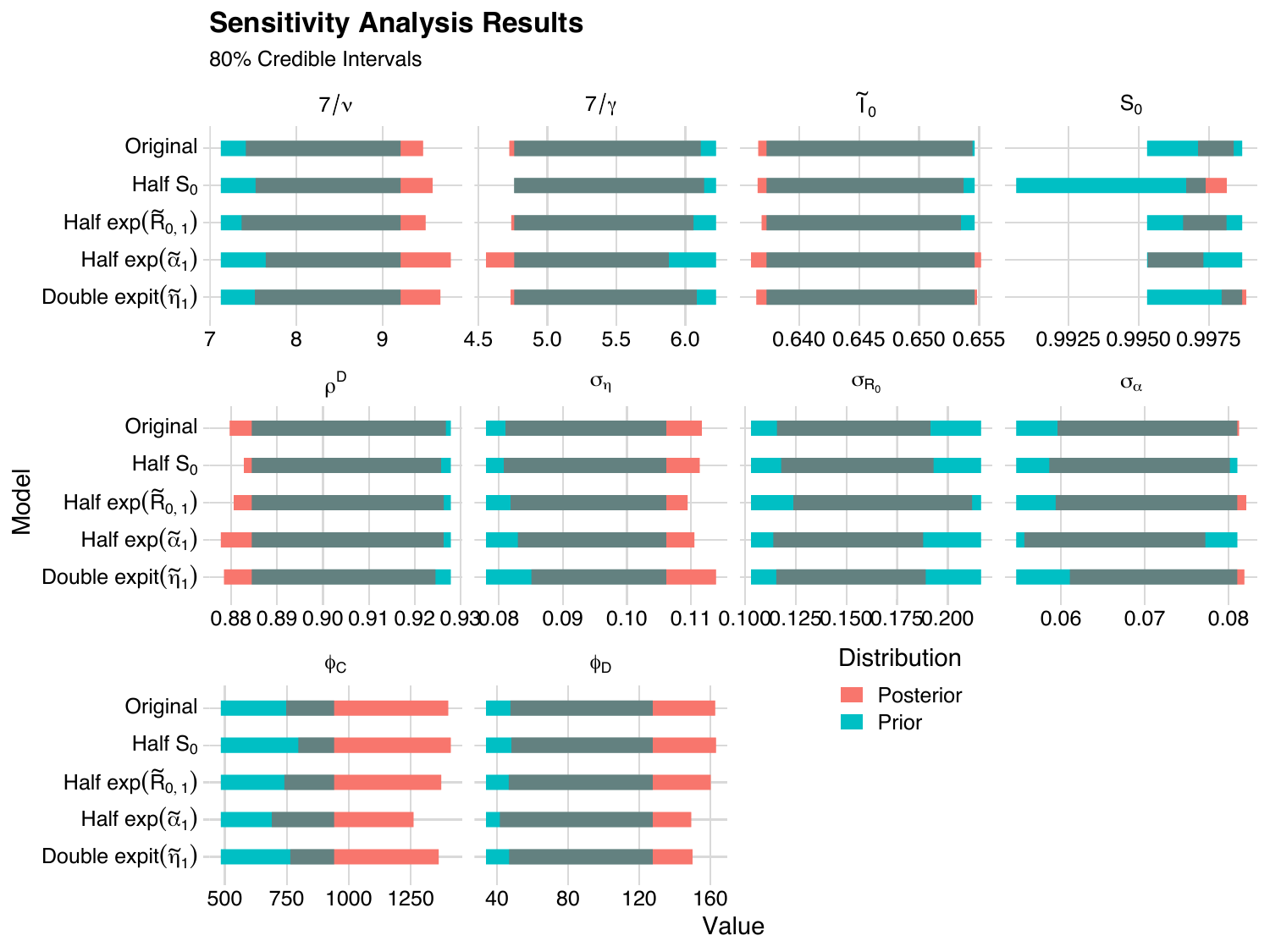}
    \caption{
    Prior and posterior 80\% credible intervals for scalar parameters from four sensitivity analyses and the original analysis.}
    \label{fig:scalar_sensitivity_plot}
\end{figure}

\begin{figure}
    \centering
    \includegraphics[width=1.0\textwidth]{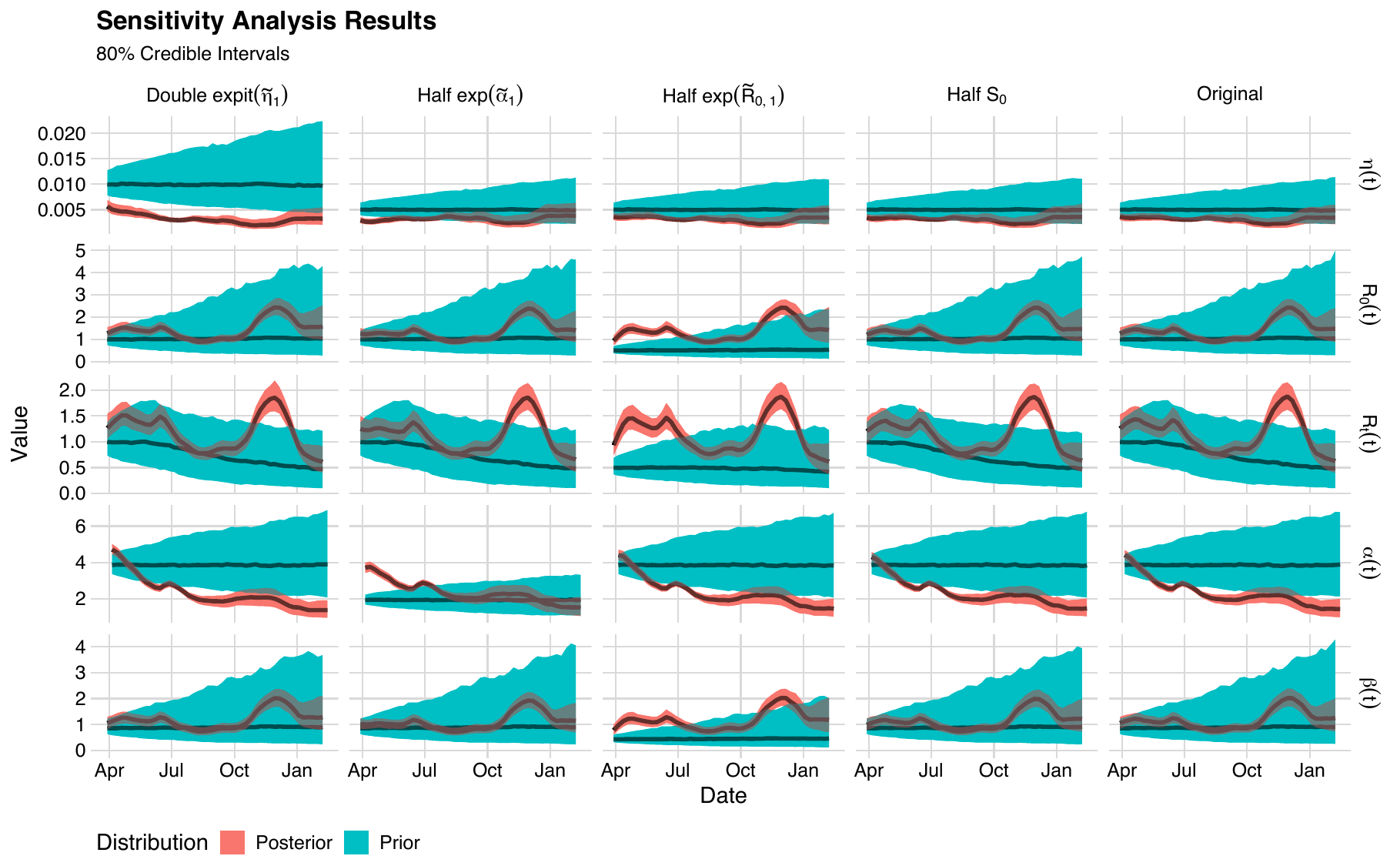}
    \caption{
    Prior and posterior 80\% credible intervals for time-varying parameters from four sensitivity analyses and the original analysis.}
    \label{fig:time_varying_sensitivity_plot}
\end{figure}

\begin{figure}
    \centering
    \includegraphics[width=1.0\textwidth]{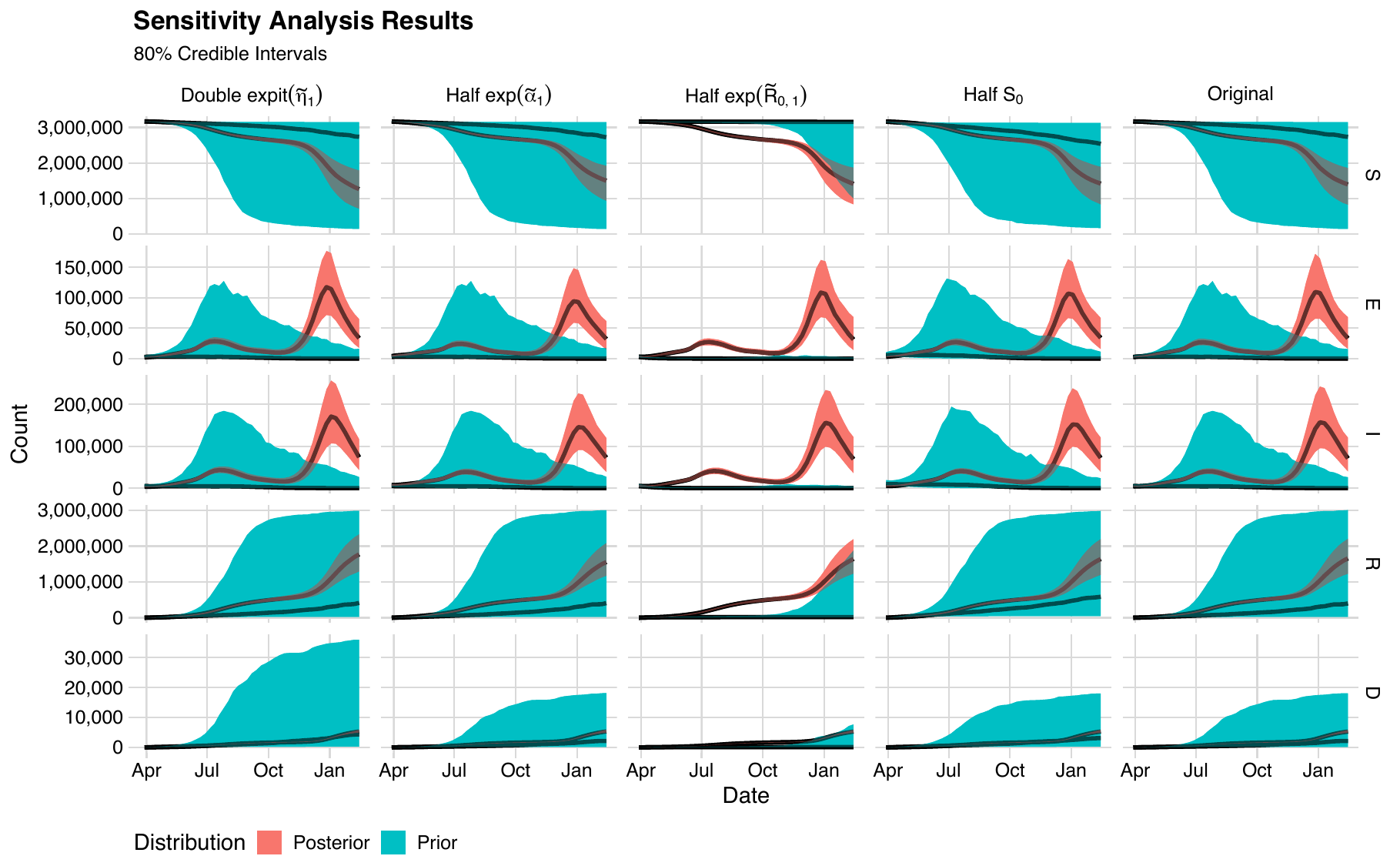}
    \caption{Prior and posterior 80\% credible intervals for time-varying parameters from four sensitivity analyses and the original analysis.}
    \label{fig:compartments_sensitivity_plot}
\end{figure}

We conducted four sensitivity analyses to see how our results change depending on the specified priors.
In each additional analysis, we change only one aspect of the model priors.
We perform one analysis where, \textit{a priori}, twice the number of people are initially infected (denoted Half \( S_0 \)), one with a lower initial basic reproduction number prior (denoted Half \( \exp\left( \tilde{R} _{0,1} \right) \), one with a higher initial infection fatality ratio prior (denoted Double \( \expit \left( \tilde{\eta}_1 \right) \), and one with a lower initial $\alpha$ prior (denoted Half \( \exp \left( \tilde{\alpha}_1 \right) \).
Precise descriptions of the priors used in the sensitivity analyses are presented in Table~\ref{table:sensitivity_priors}.
Graphical results of the sensitivity analyses are presented in Figures~\ref{fig:scalar_sensitivity_plot}--\ref{fig:compartments_sensitivity_plot}.
We find that our model is typically robust to these alternative priors, and no alternative model leads to substantively different conclusions.

\begin{table}
    \caption[Sensitivity analysis priors.]{Model parameters and their prior distributions.}
    \label{table:sensitivity_priors}
    \scriptsize\centering
    \begin{tabularx}{\textwidth}{cclclc}
	\thead{Analysis} & \thead{Parameter} & \thead{Original Prior}  &   \thead{Original Prior\\Median\\(95\% Interval)}  &   \thead{Sensitivity Prior} &   \thead{Sensitivity Prior\\Median\\(95\% Interval)}    \\ \hline
    Half \( S_0 \)  &   \( S_0 \)   &   Logit-Normal(6, 0.25) & \makecell{0.998 \\ (0.993, 0.999)}  &   Logit-Normal(5.31, 0.25) & \makecell{0.995 \\ (0.987, 0.998)}  \\
    Half \( \exp\left( \tilde{R} _{0,1} \right) \)  &   \( \exp\left( \tilde{R} _{0,1} \right) \)   &   Log-Normal(0, 0.0625) & \makecell{1.000 \\ (0.613, 1.630)} &   Log-Normal(-0.693, 0.0625) & \makecell{0.500 \\ (0.306, 0.816)} \\
    Double \( \expit \left( \tilde{\eta}_1 \right) \)  &    \( \expit \left( \tilde{\eta}_1 \right) \)  &   Logit-Normal(-5.3, 0.04) & \makecell{0.00497 \\ (0.00336, 0.00733)}  &   Logit-Normal(-4.61, 0.04) & \makecell{0.00988 \\ (0.00670, 0.01460)} \\
    Half \( \exp \left( \tilde{\alpha}_1 \right) \)  &  Half \( \exp \left( \tilde{\alpha}_1 \right) \) &   Log-Normal(1.35, 0.0121) & \makecell{3.86 \\ (3.11, 4.79)}  &   Log-Normal(0.657, 0.0121) & \makecell{1.93 \\ (1.55, 2.39)} \\
    \end{tabularx}
\end{table}

\begin{figure}
    \centering
    \includegraphics[width=1.0\textwidth]{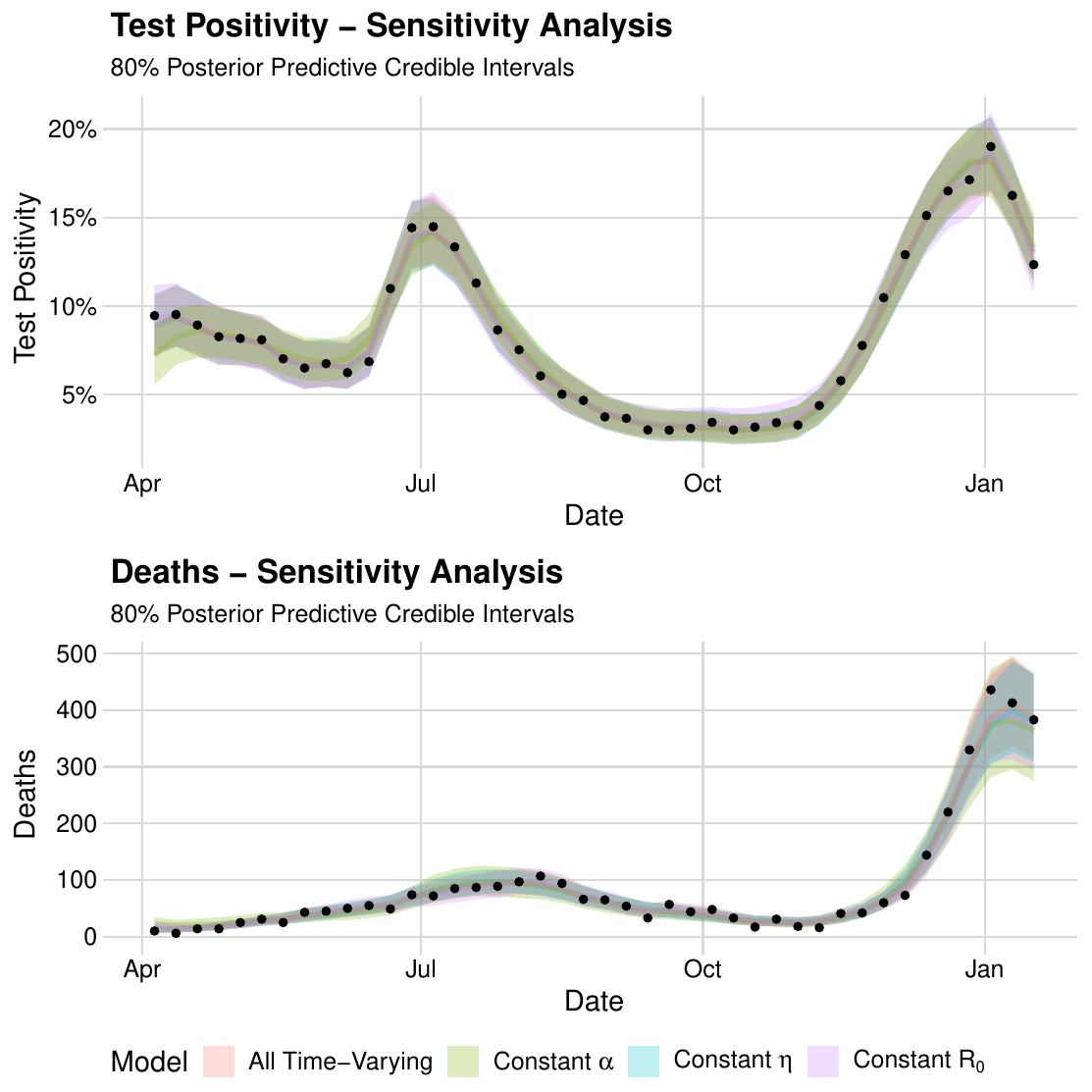}
    \caption{Posterior predictive distributions when one of the typically time-varying parameters is made to be fixed through time.}
    \label{fig:compare_constant_time_varying_posterior_predictive_plot}
\end{figure}

\begin{figure}
    \centering
    \includegraphics[width=1.0\textwidth]{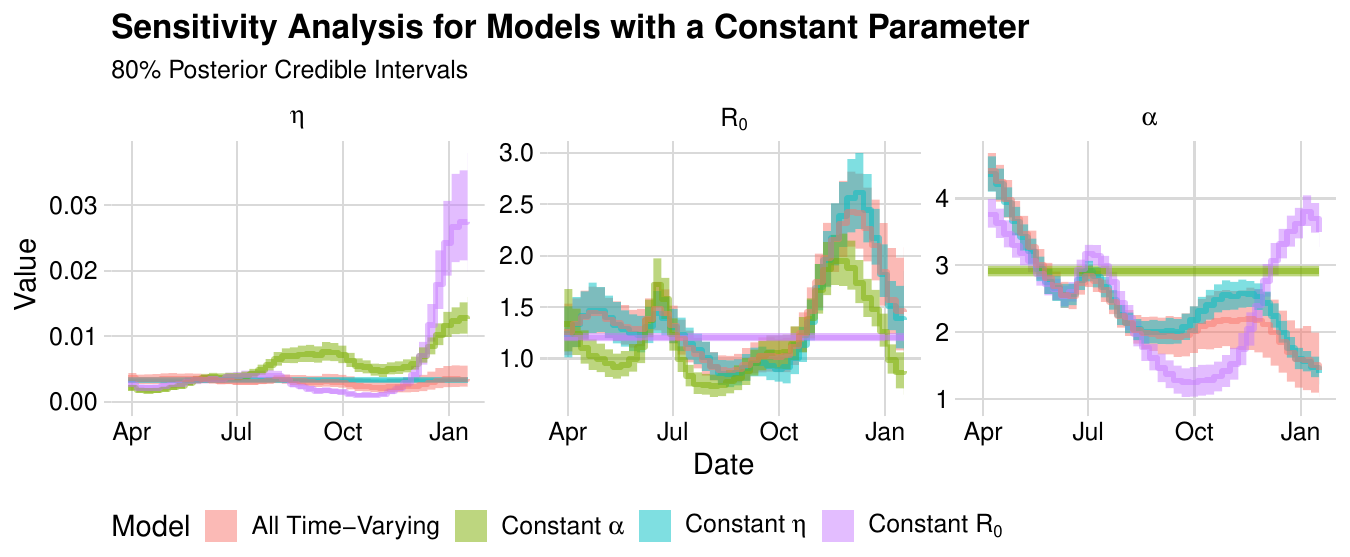}
    \caption{Posterior inference for time-varying parameters when one of the typically time-varying parameters is made to be fixed through time.}
    \label{fig:compare_constant_time_varying_inference_plot}
\end{figure}

Additionally, we perform analyses where we modify the main model to, one at a time, fix each of the time-varying parameters, \( R_0 \), \( \alpha \), and \( \eta \).
As demonstrated in Figure~\ref{fig:compare_constant_time_varying_posterior_predictive_plot}, fixing these parameters has no negative impact on the model's ability to properly fit the test positivity and death data, with each of the models exhibiting nearly identical posterior predictive distributions.
However, these modified models do lead to substantially different inferences about the time-varying parameters themselves.
This is shown in Figure~\ref{fig:compare_constant_time_varying_inference_plot}, where it appears that when one parameter is fixed, the others can become more flexible to still precisely match the observed data.
The most dramatic effect is seen when fixing \( R_0 \), which leads to drastically different inferences about \( \eta \) and \( \alpha \).
In contrast, there appears to be little impact from fixing the infection-fatality ratio, \( \eta \) as constant through time.

\section{MCMC Diagnostics}
\label{sec:convergence-diagnostics}

Convergence diagnostics are presented in Tables~\ref{tab:univariate_diagnostics} and \ref{tab:multivariate_diagnostics}, where $\hat{R}$ is the potential scale reduction factor \citep{Vehtari2021}, and ESS is the effective sample size, both as computed in the posterior R package \citep{posteriorPackage}.
All parameters show potential scale reduction factors between 1 and 1.02, providing no evidence of lack of convergence.
Additionally, all model parameters have effective sample sizes of multiple hundreds, which is sufficient for our inferences.

\begin{figure}
    \centering
    \includegraphics{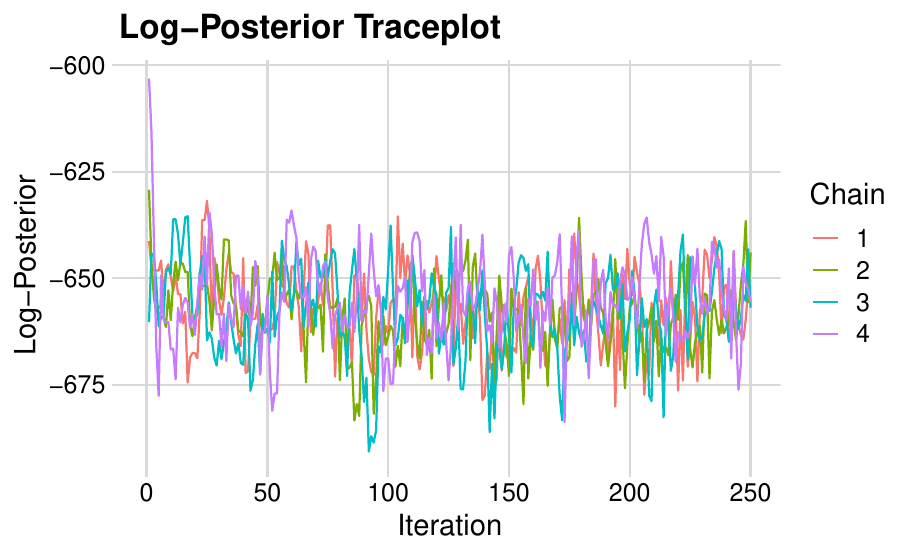}
    \caption{Trace plot of log-posterior probability for the main model fit.}
    \label{fig:lp_trace_plot}
\end{figure}

We also produce a trace plot of the log-posterior probability for each chain in Figure~\ref{fig:lp_trace_plot}, which indicates that each chain explores a region of similar probability.

\input{tables/univariate_diagnostics}
\input{tables/multivariate_diagnostics}

\end{document}

%% file: custom_commands.tex
\newcommand{\rmd}{\mathrm{d}}
\newcommand{\deriv}[2]{\frac{\rmd#1}{\rmd#2}}
\newcommand{\bX}{\mathbf{X}}
\newcommand{\bx}{\mathbf{x}}
\newcommand{\bY}{\mathbf{Y}}

\newcommand{\bZ}{\mathbf{Z}}

\newcommand{\bM}{\mathbf{M}}
\newcommand{\bN}{\mathbf{N}}

\newcommand{\etatilde}{\tilde{\eta}}
\newcommand{\alphatilde}{\tilde{\alpha}}
\newcommand{\rhoytilde}{\tilde{\rho}^Y}

\newcommand{\betatilde}{\boldsymbol{\tilde{\eta}}}
\newcommand{\balphatilde}{\boldsymbol{\tilde{\alpha}}}
\newcommand{\brtilde}{\boldsymbol{\tilde{R}_0}}
\newcommand{\brhoytilde}{\boldsymbol{\tilde{\rho}^Y}}

\newcommand{\til}[1]{\widetilde{#1}}

\newcommand{\btheta}{\boldsymbol{\theta}}

\newcommand{\Var}{\mathrm{Var}}

\DeclareMathOperator{\expit}{expit}

\newcommand{\Ie}{I_{e}}
\newcommand{\Is}{I_{p}}

\newcommand{\1}[1]{\mathds{1}\left[#1\right]}

\newcommand\bluesout{\bgroup\markoverwith{\textcolor{blue}{\rule[0.5ex]{2pt}{1.4pt}}}\ULon}

\newcommand\brownsout{\bgroup\markoverwith{\textcolor{brown}{\rule[0.5ex]{2pt}{1.4pt}}}\ULon}

%% file: tables/univariate_diagnostics.tex
\begin{table}

\caption{\label{tab:univariate_diagnostics}Convergence diagnostics for scalar parameters for the main model fit to the Orange County data set.}
\centering
\begin{tabular}[t]{lrr}
\toprule
Parameter & $\hat{R}$ & ESS\\
\midrule
$S_0$ & 1.00 & 992.14\\
$\tilde{I}_{0}$ & 1.01 & 1128.48\\
$1 / \gamma$ & 1.01 & 712.86\\
$1 / \nu$ & 1.02 & 907.22\\
$\phi_D$ & 1.00 & 876.28\\
\addlinespace
$\rho^D$ & 1.00 & 812.16\\
$\phi_C$ & 1.00 & 841.67\\
$\sigma_{R_0}$ & 1.00 & 544.21\\
$\sigma_\eta$ & 1.00 & 788.91\\
$\sigma_\alpha$ & 1.01 & 567.30\\
\bottomrule
\end{tabular}
\end{table}

%% file: tables/multivariate_diagnostics.tex
\begin{table}

\caption{\label{tab:multivariate_diagnostics}Convergence diagnostics for scalar parameters for the main model fit to the Orange County data set.}
\centering
\begin{tabular}[t]{lrrrrrr}
\toprule
Parameter & Min. $\hat{R}$ & Avg. $\hat{R}$ & Max. $\hat{R}$ & Min. ESS & Avg. ESS & Max. ESS\\
\midrule
$\exp\left(\tilde{\alpha}_t\right)$ & 1 & 1 & 1.01 & 630.78 & 827.20 & 1088.73\\
$\exp\left(\tilde{R}_{t,t}\right)$ & 1 & 1 & 1.02 & 693.98 & 977.45 & 1368.80\\
$\expit\left(\tilde{\eta}_t\right)$ & 1 & 1 & 1.02 & 563.38 & 793.53 & 1259.37\\
\bottomrule
\end{tabular}
\end{table}

%% file: oc_covid_paper.bbl
\begin{thebibliography}{45}
\newcommand{\enquote}[1]{``#1''}
\expandafter\ifx\csname natexlab\endcsname\relax\def\natexlab#1{#1}\fi

\bibitem[Anderson et~al., 2020]{anderson2020}
Anderson, S.~C., Edwards, A.~M., Yerlanov, M., Mulberry, N., Stockdale, J.~E.,
  Iyaniwura, S.~A., Falcao, R.~C., Otterstatter, M.~C., Irvine, M.~A., Janjua,
  N.~Z., Coombs, D., and Colijn, C. (2020), \enquote{Quantifying the impact of
  {COVID}-19 control measures using a {B}ayesian model of physical distancing,}
  {\em PLOS Computational Biology\/}, 16, 1--15.

\bibitem[{Andrieu} et~al., 2010]{andrieu2010particle}
{Andrieu}, C., {Doucet}, A., and {Holenstein}, R. (2010), \enquote{Particle
  {M}arkov chain {M}onte {C}arlo methods,} {\em Journal of the Royal
  Statistical Society: Series B (Statistical Methodology)\/}, 72, 269--342.

\bibitem[Bhargava et~al., 2020]{bhargava2020predictors}
Bhargava, A., Fukushima, E.~A., Levine, M., Zhao, W., Tanveer, F., Szpunar,
  S.~M., and Saravolatz, L. (2020), \enquote{Predictors for severe {COVID}-19
  infection,} {\em Clinical Infectious Diseases\/}, 71, 1962--1968.

\bibitem[Bosse et~al., 2022]{scoringutils}
Bosse, N.~I., Gruson, H., Cori, A., van Leeuwen, E., Funk, S., and Abbott, S.
  (2022), \enquote{Evaluating forecasts with scoringutils in {R},} {\em arXiv
  preprint arXiv:2205.07090\/}.

\bibitem[{Bret{\'o}} et~al., 2009]{breto2009time}
{Bret{\'o}}, C., {He}, D., {Ionides}, E., and {King}, A. (2009), \enquote{Time
  series analysis via mechanistic models,} {\em The Annals of Applied
  Statistics\/}, 3, 319--348.

\bibitem[{Bret{\'o}} and {Ionides}, 2011]{breto2011compound}
{Bret{\'o}}, C. and {Ionides}, E. (2011), \enquote{Compound {M}arkov counting
  processes and their applications to modeling infinitesimally over--dispersed
  systems,} {\em Stochastic Processes and their Applications\/}, 121,
  2571--2591.

\bibitem[Bruckner et~al., 2021]{Bruckner2021}
Bruckner, T.~A., Parker, D.~M., Bartell, S.~M., Vieira, V.~M., Khan, S.,
  Noymer, A., Drum, E., Albala, B., Zahn, M., and Boden-Albala, B. (2021),
  \enquote{Estimated seroprevalence of {SARS-CoV-2} antibodies among adults in
  {O}range {C}ounty, {C}alifornia,} {\em Scientific Reports\/}, 11, 3081.

\bibitem[Byrne et~al., 2020]{Byrnee039856}
Byrne, A.~W., McEvoy, D., Collins, A.~B., Hunt, K., Casey, M., Barber, A.,
  Butler, F., Griffin, J., Lane, E.~A., McAloon, C., O{\textquoteright}Brien,
  K., Wall, P., Walsh, K.~A., and More, S.~J. (2020), \enquote{Inferred
  duration of infectious period of {SARS-CoV-2}: rapid scoping review and
  analysis of available evidence for asymptomatic and symptomatic {COVID-19}
  cases,} {\em BMJ Open\/}, 10.

\bibitem[Bürkner et~al., 2022]{posteriorPackage}
Bürkner, P.-C., Gabry, J., Kay, M., and Vehtari, A. (2022),
  \enquote{posterior: Tools for working with posterior distributions,} R
  package version 1.3.1.

\bibitem[{Cauchemez} and {Ferguson}, 2008]{cauchemez2008likelihood}
{Cauchemez}, S. and {Ferguson}, N. (2008), \enquote{Likelihood-based estimation
  of continuous-time epidemic models from time-series data: application to
  measles transmission in {L}ondon,} {\em Journal of the Royal Society
  Interface\/}, 5, 885--897.

\bibitem[Cummings et~al., 2020]{Cummings2020}
Cummings, M.~J., Baldwin, M.~R., Abrams, D., Jacobson, S.~D., Meyer, B.~J.,
  Balough, E.~M., Aaron, J.~G., Claassen, J., Rabbani, L.~E., Hastie, J.,
  et~al. (2020), \enquote{Epidemiology, clinical course, and outcomes of
  critically ill adults with {COVID}-19 in {N}ew {Y}ork {C}ity: a prospective
  cohort study,} {\em The Lancet\/}, 395, 1763--1770.

\bibitem[Davies et~al., 2021]{Davieseabg2021}
Davies, N.~G., Abbott, S., Barnard, R.~C., Jarvis, C.~I., Kucharski, A.~J.,
  Munday, J.~D., Pearson, C. A.~B., Russell, T.~W., Tully, D.~C., Washburne,
  A.~D., Wenseleers, T., Gimma, A., Waites, W., Wong, K. L.~M., van Zandvoort,
  K., Silverman, J.~D., Group, C. C.-.~W., Consortium, C.-. G. U. C.-U.,
  Diaz-Ordaz, K., Keogh, R., Eggo, R.~M., Funk, S., Jit, M., Atkins, K.~E., and
  Edmunds, W.~J. (2021), \enquote{Estimated transmissibility and impact of
  {SARS-CoV-2} lineage {B.1.1.7} in {E}ngland,} {\em Science\/}, 372, ISSN
  0036-8075.

\bibitem[Davies et~al., 2020]{davies2020}
Davies, N.~G., Kucharski, A.~J., Eggo, R.~M., Gimma, A., Edmunds, W.~J.,
  Jombart, T., O'Reilly, K., Endo, A., Hellewell, J., Nightingale, E.~S.,
  et~al. (2020), \enquote{Effects of non-pharmaceutical interventions on
  {COVID}-19 cases, deaths, and demand for hospital services in the {UK}: a
  modelling study,} {\em The Lancet Public Health\/}, 5, e375--e385.

\bibitem[Dukic et~al., 2012]{dukic2012tracking}
Dukic, V., Lopes, H., and Polson, N. (2012), \enquote{Tracking epidemics with
  {G}oogle flu trends data and a state-space {SEIR} model,} {\em Journal of the
  American Statistical Association\/}, 107, 1410--1426.

\bibitem[{Ferguson} and {et. al}, 2020]{ferguson2020report9}
{Ferguson}, N. and {et. al} (2020), \enquote{Report 9: Impact of
  non-pharmaceutical interventions ({NPI}s) to reduce {COVID}-19 mortality and
  healthcare demand,} {\em MRC Centre for Global Infectious Disease Analysis
  Reports\/}, accessed: 2020-06-19.

\bibitem[Fintzi et~al., 2022]{fintzi2020linear}
Fintzi, J., Wakefield, J., and Minin, V.~N. (2022), \enquote{A linear noise
  approximation for stochastic epidemic models fit to partially observed
  incidence counts,} {\em Biometrics\/}, 78, 1530--1541.

\bibitem[Ge et~al., 2018]{turing}
Ge, H., Xu, K., and Ghahramani, Z. (2018), \enquote{Turing: A language for
  flexible probabilistic inference,} in {\em Proceedings of the Twenty-First
  International Conference on Artificial Intelligence and Statistics\/}, edited
  by A.~Storkey and F.~Perez-Cruz, volume~84 of {\em Proceedings of Machine
  Learning Research\/}, pages 1682--1690, PMLR.

\bibitem[Grint et~al., 2021]{grint2021severity}
Grint, D.~J., Wing, K., Houlihan, C., Gibbs, H.~P., Evans, S.~J., Williamson,
  E., McDonald, H.~I., Bhaskaran, K., Evans, D., Walker, A.~J., et~al. (2021),
  \enquote{Severity of {SARS-CoV-2} alpha variant ({B}.1.1.7) in {E}ngland,}
  {\em Clinical Infectious Diseases\/}, in press.

\bibitem[Hoffman and Gelman, 2014]{NUTS}
Hoffman, M.~D. and Gelman, A. (2014), \enquote{The {N}o-{U}-{T}urn sampler:
  Adaptively setting path lengths in {H}amiltonian {M}onte {C}arlo,} {\em
  Journal of Machine Learning Research\/}, 15, 1593--1623.

\bibitem[{H{\"o}hle} and {an der Heiden}, 2014]{hohle2014bayesian}
{H{\"o}hle}, M. and {an der Heiden}, M. (2014), \enquote{Bayesian nowcasting
  during the {STEC O104: H4} outbreak in {G}ermany, 2011,} {\em Biometrics\/},
  70, 993--1002.

\bibitem[Irons and Raftery, 2021]{irons2021estimating}
Irons, N.~J. and Raftery, A.~E. (2021), \enquote{Estimating {SARS-CoV}-2
  infections from deaths, confirmed cases, tests, and random surveys,} {\em
  Proceedings of the National Academy of Sciences\/}, 118, e2103272118.

\bibitem[Jewell et~al., 2021]{miller2020}
Jewell, S., Futoma, J., Hannah, L., Miller, A.~C., Foti, N.~J., and Fox, E.~B.
  (2021), \enquote{It's complicated: characterizing the time-varying
  relationship between cell phone mobility and {COVID-19} spread in the {US},}
  {\em npj Digital Medicine\/}, 4, 152, ISSN 2398-6352.

\bibitem[Jordan et~al., 2019]{scoringRules}
Jordan, A., Krüger, F., and Lerch, S. (2019), \enquote{Evaluating
  probabilistic forecasts with {scoringRules},} {\em Journal of Statistical
  Software\/}, 90, 1–37.

\bibitem[Kim et~al., 2021]{kim2020risk}
Kim, L., Garg, S., O?Halloran, A., Whitaker, M., Pham, H., Anderson, E.~J.,
  Armistead, I., Bennett, N.~M., Billing, L., Como-Sabetti, K., et~al. (2021),
  \enquote{Risk factors for intensive care unit admission and in-hospital
  mortality among hospitalized adults identified through the {US} coronavirus
  disease 2019 ({COVID}-19)-associated hospitalization surveillance network
  ({COVID-NET}),} {\em Clinical Infectious Diseases\/}, 72, e206--e214.

\bibitem[Knock et~al., 2021]{Knockeabg2021}
Knock, E.~S., Whittles, L.~K., Lees, J.~A., Perez-Guzman, P.~N., Verity, R.,
  FitzJohn, R.~G., Gaythorpe, K. A.~M., Imai, N., Hinsley, W., Okell, L.~C.,
  Rosello, A., Kantas, N., Walters, C.~E., Bhatia, S., Watson, O.~J.,
  Whittaker, C., Cattarino, L., Boonyasiri, A., Djaafara, B.~A., Fraser, K.,
  Fu, H., Wang, H., Xi, X., Donnelly, C.~A., Jauneikaite, E., Laydon, D.~J.,
  White, P.~J., Ghani, A.~C., Ferguson, N.~M., Cori, A., and Baguelin, M.
  (2021), \enquote{Key epidemiological drivers and impact of interventions in
  the 2020 {SARS-CoV}-2 epidemic in {E}ngland,} {\em Science Translational
  Medicine\/}, 13, eabg4262.

\bibitem[{Lekone} and {Finkenst{\"a}dt}, 2006]{lekone2006statistical}
{Lekone}, P. and {Finkenst{\"a}dt}, B. (2006), \enquote{Statistical inference
  in a stochastic epidemic {SEIR} model with control intervention: {E}bola as a
  case study,} {\em Biometrics\/}, 62, 1170--1177.

\bibitem[{Li} and {Brauer}, 2008]{li2008continuous}
{Li}, J. and {Brauer}, F. (2008), \enquote{Continuous-time age-structured
  models in population dynamics and epidemiology,} in {\em Mathematical
  Epidemiology\/}, edited by F.~Brauer, P.~van~den Driessche, and J.~Wu,
  chapter~9, pages 205--227, Springer.

\bibitem[Matheson and Winkler, 1976]{CRPS}
Matheson, J.~E. and Winkler, R.~L. (1976), \enquote{Scoring rules for
  continuous probability distributions,} {\em Management Science\/}, 22,
  1087--1096.

\bibitem[Morozova et~al., 2021]{morozova2021one}
Morozova, O., Li, Z.~R., and Crawford, F.~W. (2021), \enquote{One year of
  modeling and forecasting {COVID}-19 transmission to support policymakers in
  {C}onnecticut,} {\em Scientific Reports\/}, 13, 20271.

\bibitem[Nguyen-Van-Yen et~al., 2021]{nguyen2021stochastic}
Nguyen-Van-Yen, B., {Del Moral}, P., and Cazelles, B. (2021),
  \enquote{Stochastic epidemic models inference and diagnosis with {P}oisson
  random measure data augmentation,} {\em Mathematical Biosciences\/}, 335,
  108583.

\bibitem[Pei et~al., 2021]{pei2021burden}
Pei, S., Yamana, T.~K., Kandula, S., Galanti, M., and Shaman, J. (2021),
  \enquote{Burden and characteristics of {COVID}-19 in the {U}nited {S}tates
  during 2020,} {\em Nature\/}, pages 1--18.

\bibitem[Petrilli et~al., 2020]{petrilli2020factors}
Petrilli, C.~M., Jones, S.~A., Yang, J., Rajagopalan, H., O'Donnell, L.,
  Chernyak, Y., Tobin, K.~A., Cerfolio, R.~J., Francois, F., and Horwitz, L.~I.
  (2020), \enquote{Factors associated with hospital admission and critical
  illness among 5279 people with coronavirus disease 2019 in {N}ew {Y}ork
  {C}ity: prospective cohort study,} {\em BMJ\/}, 369.

\bibitem[{Pooley} et~al., 2015]{pooley2015using}
{Pooley}, C., {Bishop}, S., and {Marion}, G. (2015), \enquote{Using model-based
  proposals for fast parameter inference on discrete state space,
  continuous-time {M}arkov processes,} {\em Journal of The Royal Society
  Interface\/}, 12, 20150225.

\bibitem[{Prem} and {et. al}, 2020]{prem2020effect}
{Prem}, K. and {et. al} (2020), \enquote{The effect of control strategies to
  reduce social mixing on outcomes of the {COVID}-19 epidemic in {W}uhan,
  {C}hina: a modelling study,} {\em The Lancet Public Health\/}, 5, e261--e270.

\bibitem[Scott et~al., 2020]{epidemia}
Scott, J.~A., Gandy, A., Mishra, S., Unwin, J., Flaxman, S., and Bhatt, S.
  (2020), \enquote{epidemia: Modeling of epidemics using hierarchical
  {B}ayesian models,} R package version 1.0.0.

\bibitem[Song et~al., 2020]{Song2020}
Song, J.-W., Zhang, C., Fan, X., Meng, F.-P., Xu, Z., Xia, P., Cao, W.-J.,
  Yang, T., Dai, X.-P., Wang, S.-Y., et~al. (2020), \enquote{Immunological and
  inflammatory profiles in mild and severe cases of {COVID}-19,} {\em Nature
  Communications\/}, 11, 1--10.

\bibitem[{Stan Development Team}, 2020]{rstan}
{Stan Development Team} (2020), \enquote{{RStan}: the {R} interface to {Stan},}
  R package version 2.21.2.

\bibitem[Stokes et~al., 2021]{stokes2020assessing}
Stokes, A.~C., Lundberg, D.~J., Elo, I.~T., Hempstead, K., Bor, J., and
  Preston, S.~H. (2021), \enquote{{COVID}-19 and excess mortality in the
  {U}nited {S}tates: A county-level analysis,} {\em PLOS Medicine\/}, 18,
  1--18.

\bibitem[Stoner and Economou, 2020]{stoner2019multivariate}
Stoner, O. and Economou, T. (2020), \enquote{Multivariate hierarchical
  frameworks for modeling delayed reporting in count data,} {\em Biometrics\/},
  76, 789--798.

\bibitem[{United States Census Bureau}, 2020]{orangecensus}
{United States Census Bureau} (2020), \enquote{Quick facts: {O}range {C}ounty,
  {C}alifornia,}
  \url{https://www.census.gov/quickfacts/orangecountycalifornia}, accessed:
  2020-09-05.

\bibitem[{Van den Driessche}, 2008]{van2008spatial}
{Van den Driessche}, P. (2008), \enquote{Spatial structure: Patch models,} in
  {\em Mathematical Epidemiology\/}, edited by F.~Brauer, P.~van~den Driessche,
  and J.~Wu, chapter~7, pages 179--189, Springer.

\bibitem[Vehtari et~al., 2021]{Vehtari2021}
Vehtari, A., Gelman, A., Simpson, D., Carpenter, B., and B{\"u}rkner, P.-C.
  (2021), \enquote{{Rank-Normalization, Folding, and Localization: An Improved
  $\widehat{R}$ for Assessing Convergence of MCMC (with Discussion)},} {\em
  Bayesian Analysis\/}, 16, 667 -- 718.

\bibitem[WHO, 2021]{whocoronavirustrans}
WHO (2021), \enquote{Word {H}ealth {O}rganization q\&a: {C}oronavirus disease
  ({COVID}-19): {H}ow is it transmitted?}
  \url{https://www.who.int/emergencies/diseases/novel-coronavirus-2019/question-and-answers-hub/q-a-detail/coronavirus-disease-covid-19-how-is-it-transmitted},
  accessed: 2021-09-12.

\bibitem[Wu and McGoogan, 2020]{Wu2020CDC}
Wu, Z. and McGoogan, J.~M. (2020), \enquote{Characteristics of and important
  lessons from the coronavirus disease 2019 ({COVID}-19) outbreak in {C}hina:
  summary of a report of 72 314 cases from the {C}hinese {C}enter for {D}isease
  {C}ontrol and {P}revention,} {\em JAMA\/}, 323, 1239--1242.

\bibitem[Xin et~al., 2021]{Xin2021}
Xin, H., Li, Y., Wu, P., Li, Z., Lau, E. H.~Y., Qin, Y., Wang, L., Cowling,
  B.~J., Tsang, T.~K., and Li, Z. (2021), \enquote{{Estimating the Latent
  Period of Coronavirus Disease 2019 (COVID-19)},} {\em Clinical Infectious
  Diseases\/}, 74, 1678--1681, ISSN 1058-4838.

\end{thebibliography}
